\documentclass[pdftex,12pt,preprint]{aastex}

\usepackage{natbib}
\usepackage[dvips]{color}

\usepackage{graphics}
\usepackage{epsfig}
\usepackage{amsmath}
\usepackage{amssymb}
\usepackage{pdflscape}
\usepackage{color}

\newcommand{\re}{$R_{\oplus}$}

\shorttitle{Post- Giant Impact Earths}
\shortauthors{Lupu et al.}

\begin{document}

\title{THE ATMOSPHERES OF EARTH-LIKE PLANETS AFTER GIANT IMPACT EVENTS}

\author{R. E. Lupu,$^{1\ast}$ Kevin Zahnle,$^{2}$ Mark\ S. Marley,$^{2}$ Laura Schaefer,$^{3}$ Bruce Fegley,$^{4}$ Caroline Morley,$^{5}$ Kerri Cahoy,$^{6}$ 
Richard Freedman,$^{1}$ Jonathan J. Fortney,$^{5}$ \\
\vspace{0.2in}
\footnotesize{$^{1}$ SETI Institute / NASA Ames Research Center, Moffet Field, CA 94035, USA; Roxana.E.Lupu@nasa.gov}\\
$^{2}$ NASA Ames Research Center, Moffet Field, CA 94035, USA\\
$^{3}$ Harvard/Smithsonian Center for Astrophysics, Cambridge, MA 02138, USA\\
$^{4}$ Planetary Chemistry Laboratory, Dept. of Earth \& Planetary Sciences and McDonnell Center for the Space Sciences, Washington University, St Louis, MO 63130 USA\\
$^{5}$ Department of Astronomy and Astrophysics, University of California,
Santa Cruz, CA 95064, USA\\
$^{6}$ Massachusetts Institute of Technology, Cambridge, MA 02139, USA}

\pagestyle{myheadings}
\markright{Draft \today}
\markright{\today}

\begin{abstract}

 It is now understood that the accretion of terrestrial planets naturally involves giant collisions, the moon-forming impact being a well known example. In the aftermath of such collisions the surface of the surviving planet is very hot and potentially detectable.  Here we explore the atmospheric chemistry, photochemistry, and spectral signatures of post-giant-impact terrestrial planets enveloped by thick atmospheres consisting predominantly of $\rm CO_2$, and $\rm H_2O$. The atmospheric chemistry and structure are computed self-consistently for atmospheres in equilibrium with hot surfaces with composition reflecting either the bulk silicate Earth (which includes the crust, mantle, atmosphere and oceans) or Earth's continental crust. We account for all major molecular and atomic opacity sources including collision-induced absorption. We find that these atmospheres are dominated by H$_2$O and CO$_2$, while the formation of CH$_4$, and NH$_3$ is quenched due to short dynamical timescales. Other important constituents are HF, HCl, NaCl, and SO$_2$. These are apparent in the emerging spectra, and can be indicative that an impact has occurred. The use of comprehensive opacities results in spectra that are a factor of 2 lower in surface brightness in the spectral windows than predicted by previous models. The estimated luminosities show that the hottest post-giant-impact planets will be detectable with near-infrared coronagraphs on the planned $30\,\rm m$-class telescopes. The $1-4$ $\mu$m will be most favorable for such detections, offering bright features and better contrast between the planet and a potential debris disk. We derive cooling timescales on the order of 10$^{5-6}$ years, based on the modeled effective temperatures. This leads to the possibility of discovering tens of such planets in future surveys.

\end{abstract}


\section{INTRODUCTION\label{intro}}

The final assembly of terrestrial planets is now universally thought to have occurred through a series of giant impacts -- essentially collisions between planets -- spread out over some 100 million years \citep{Cameron:1976,Wetherill:1985,Lissauer:1993,Chambers:2004,Raymond:2004}. The most famous of these is Earth's own Moon-forming impact \citep{Hartmann:1986,Benz:1986,Canup:2000, Canup:2001,Agnor:2004,Canup:2004}. It takes at least 10 collisions between planets to make a Venus and an Earth, as not every collision results in a merger. In the aftermath of one of these collisions the surviving planet is hot, and can remain hot for millions of years \citep{Zahnle:2006,Zahnle:2007}.
During this phase of accretion a proto-terrestrial planet may have a dense steam atmosphere \citep[e.g.,][]{Abe:1988,Matsui:1986,Zahnle:1988,Hashimoto:2007,Schaefer:2010}. Eventually the atmosphere cools and water vapor condenses into clouds. How long the stricken planet remains hot depends on the size of the collision and the nature of the planet's atmosphere. While hot, the planet will be bright, especially in the near IR, where spectral windows reveal hotter atmospheric depths. 

Such young, post-giant impact terrestrial planets will be far brighter and easier to detect around nearby young stars than old, cold terrestrial planets \citep{Stern:1994,Miller-Ricci:2009}. Furthermore, the time period after the last giant impact sets the boundary condition for the subsequent thermal evolution of Earth. Whether a terrestrial planet retains water or enters into a runaway greenhouse ultimately depends upon the conditions after the last giant impact \citep{Hamano:2013}.  The classic early studies of runaway greenhouse atmospheres \citep[e.g.,][]{Ingersoll:1969} used a grey approximation for water opacity and neglected other opacity sources. Second generation models that greatly improved on the radiative transfer in runaway and near-runaway greenhouse atmospheres were developed independently by \citet{Kasting:1988} and \citet{Abe:1988}. These models, state-of-the-art in their day, used multiple bands of $\rm H_2O$ and $\rm CO_2$, but the description of hot bands was necessarily crude given what was known at the time, and these models completely neglected other opacity sources, some of which (for example the alkali metals) are now known (thanks to brown dwarf science, e.g., \citet{Burrows:2000}) to be of first-order importance.

The composition of a post-impact atmosphere is unlikely to be a pure mixture of $\rm H_2O$ and $\rm CO_2$. At a minimum volatile species will evaporate and the atmosphere will equilibrate with the surface. Thus the atmospheric composition depends upon the surface composition. Whether or not the surface is oxidized (Fe$^{3+}$ minerals) vs. reduced (Fe$^{2+}$ minerals), for example, will control the oxidation state of carbon compounds in the atmosphere. Thus any model of the atmospheric structure of a post impact world must consider a range of surface compositions.

Radiometric dating shows that the Earth's continental crust formed very early in the history of the Earth. As noted by \citet{Fegley:2012}, coupled modeling of the short-lived $^{182}$Hf$-^{182}$W and $^{146}$Sm$-^{142}$Nd systems by \citet{Moynier:2010} shows Earth's crust formed 38$-$75 million years after the formation of the solar system. The oldest dated crustal samples are detrital zircons from Jack Hills, Australia, which are up to 4,404 million years old \citep{Wilde:2001,Harrison:2005}. These zircons show that continental crust existed 164 million years after formation of the solar system. Thus it is likely that Earth-like rocky exoplanets also formed felsic crusts very early in their evolution.  However the alternative case, that the surface of a post-impact world does not yet reflect the formation of continents and is composed mostly of mafic silicates, like the bulk silicate Earth, should also be considered.

In this paper we report on our calculation of the atmospheric thermal structure, equilibrium and disequilibrium chemistry, emergent spectra, and thermal evolution of post-giant impact terrestrial planets.  Based on these results we also discuss the detectability of such worlds and the spectral markers that can discriminate between different surface chemistries.  The framework we present is relevant to a variety of problems including the runaway greenhouse and the early evolution of the Earth-Moon system. We necessarily neglect some processes that might be important, such as cloud formation and a self-consistent photochemical model, which are deferred to a future paper.

\section{MODELING\label{models}}

To model the atmospheric thermal structure of post-impact worlds we construct a suite of radiative-convective
equilibrium models that are consistent with a specified range of surface temperatures. For each model we specify the incident flux, surface gravity and pressure, 
and  surface chemical composition. We note that the impacted planet cools over a timescale of 10$^{5}$-10$^{6}$ years \citep{Zahnle:2007} while the timescale for atmospheric adjustment to radiative and convective equilibrium is far shorter, in the range of a day or so. Thus even though the atmosphere will cool through time, the thermal structure for a given surface temperature is well defined. This is the same procedure that is commonly used for evolution calculations for gas giants where a single atmospheric boundary is associated with a cooling planet or brown dwarf.

For a specified surface temperature we solve for the radiative-convective $T-p$ profile taking into account the heating from the surface and the incoming solar radiation, ensuring energy balance throughout the atmosphere. The atmosphere code we employ iteratively solves for radiative-convective equilibrium by adjusting the size of the convection zone until the lapse rate everywhere in the convective region is sub-adiabatic. This code was originally developed for modeling Titan's atmosphere \citep{McKay:1989}, and has been extensively modified and applied to the study of brown dwarfs \citep[e.g.,][]{Marley:1996,Cushing:2008,Saumon:2008}, and solar and extrasolar giant planets \citep{Marley:1999,Marley:1999u,Fortney:2008a,Fortney:2008b}. When fully converged the radiative layers are typically in full radiative energy balance to about 1 part in $10^5$.
Our solution achieves this energy balance by using a matrix method that progressively adjusts an initial guess at the $T-p$ profile  \citep{Pollack:1973,McKay:1989} rather than by time stepping \citep[e.g., ][]{Gierasch:1968,Pavlov:2000}. Both approaches are intrinsically quasi-static, with the matrix method being more efficient \citep[e.g., ][]{Ramanathan:1978}.

To derive the atmospheric composition we rely on thermochemical equilibrium calculations for ``Earth-like'' rocky planets that are heated to high temperatures \citep{Schaefer:2012}. The chemistry calculations are done for a grid with temperatures from 300$-$4000~K and one microbar$-$300~bar pressures, using a Gibbs energy minimization code (see \citet{Zeggern:1970}). Approximately $\sim$800 compounds (solid, liquid, gas) of the elements are included in the calculations. We use two reasonable representative compositions of the silicate portion of rocky planets: (1) the Earth's continental crust (CC, \citet{Wedepohl:1995}) and (2) the bulk silicate Earth (BSE, \citet{Kargel:1993}). 
These two compositions allow us to examine differences in atmospheres formed by outgassing of silica-rich (felsic) rocks $-$ like the Earth's continental crust $-$ and MgO- and FeO-rich (mafic) rocks $-$ like the BSE. These compositions describe the class of rocky planets around stars of solar metallicity, with a differentiated Fe metal and FeS core. For reference, the BSE is the composition of the silicate portion of the Earth that evolved into the present-day mantle, crust, oceans, and atmosphere. The mantle comprises 99.4\% by mass of the BSE, and thus the BSE composition is similar to that of Earth's mantle \citep{Schaefer:2012}. The most important compounds considered in the calculations of CC and BSE compositions are listed in Table~\ref{tab:comp}. The breakdown in terms of the most abundant elements for the two compositions is listed in the same table.

For both compositions, we generate a grid of atmosphere models for surface pressures $p_{\rm surf}$ of 10 and $100\,\rm bar$, surface gravities $g = 10$ and $30\,\rm m\, s^{-2}$, and spanning surface temperatures from 1000 to 2200~K in steps of 200~K. The atmospheres are heated both from below, given the specified surface temperature, and from above by the incident visible flux from the sun at 1 AU. Other stellar types, including a potential faint young Sun (appropriate for the Moon-forming impact), are not considered in this work. Insolation is consistenlty treated as part of the energy balance in the planet's atmosphere, and it affects the resulting $T-p$ profile. As shown in section \ref{sec:pt}, in the runaway greenhouse limit the incident stellar flux sets the lowest surface temperature the planet can attain. Outside this limit, we check the effect of the incident radiation on the effective temperature of the planet by varying the star-planet distance, and find it to be small (see Figure \ref{sun_dist}).

Although condensation is accounted for in the chemical equilibrium abundances and gas opacities, we neglect here the opacity of cloud forming species and we do not compute a wet adiabat as water condensation generally does not occur in the model tropospheres. In our models the most important condensing species is NaCl. Most Fe metal remains in the magma and is never vaporized in our models \citep{Schaefer:2012}, and therefore Fe metal condensation is not an issue. We estimate that the effects of the latent heat released by the condensation of NaCl on the adiabatic lapse rate is negligible compared to the gas C$_p$. For a condensing species, the effect of the latent heat on the lapse rate is given by the term $LX_i(L/$R$T^2-1/T)$, where $L$ is the latent heat, R is the ideal gas constant, and $X_i$ is the mole fraction of the condensing species \citep{Lewis:1969}. At 1600 K, the enthalpy of reaction for NaCl (gas) = NaCl (liquid) is -170,217 J/mol, and the corresponding heat effect will be 1.25 J/mol K, where $X_i$ for NaCl is 0.001. This effect from the condensing NaCl is only 2.4\% of the  C$_p$ of the gas itself, which is about 53.47 J/mol K (for a 50:50 CO$_2$-H$_2$O mixture). We conclude that the departures from the dry adiabat are insignificant for the cases presented in this paper.

The radiative transfer equations for incoming solar radiation are solved using the delta-discrete ordinates method of \citet{Toon:1989} while thermal radiative transfer is computed using the two-stream source function method, also of \citet{Toon:1989}.  With a few exceptions our model planets are assumed to be located at 1~AU from the sun and a current-day solar spectrum is used as the external radiation flux.  

The radiative transfer calculation treats opacity with the correlated-k technique \citep{Goody:1989} for computational speed, with 196 wavelength intervals over the whole spectrum. The correlated-k coefficients are derived using 33 individual molecular opacity sources plus alkali metals, as listed in Table~\ref{tab:opac}. Hot line lists are used for every species for which they are available as described in  \citet{Freedman:2008}. In addition we extend the opacities mentioned in Freedman et al.\ using additional HITRAN data \citep{Rothman:2009}. For completeness, the literature resources for the molecular opacities used in this paper are also summarized in Table~\ref{tab:opac}. The listed H$_2$O and CO$_2$ databases are in agreement with HITEMP 2010 \citep{Rothman:2010} for the range of temperatures and pressures relevant to this paper. Our molecular and atomic opacity database is adequate for temperatures up to a few thousand degrees, as checked against the latest laboratory experiments, and is used as a validation tool for state-of-the-art theoretical calculations. It has also been used for modeling the spectra of brown dwarfs with effective temperatures between 500 and 2400 K, with excellent agreement with the observations \citep[e.g.][]{Cushing:2008,Saumon:2008}. 

The opacities for each molecule are pre-computed at high spectral resolution on a 1060-point pressure-temperature grid, from $75-4000$~K and $10^{-6}-300$~bar. The molecular and alkali chemical equilibrium abundances for each of the two compositions are interpolated on the same 1060-point $(p,T)$ grid, and used to pre-mix the opacities before deriving the k-coefficients. The abundances were extrapolated below 300~K by requiring that the partial pressure of water vapor not exceed the saturated vapor pressure at the given temperature. 

In addition to the opacity sources included in the k-coefficients, the radiative transfer takes into account the opacity arising from collisionally-induced absorption by H$_{2}$ \citep[and references therein]{Freedman:2008}, as well as other CIA opacity for molecules recently available through the HITRAN database \citep[and references therein]{Richard:2012}, i.e. CH$_4$, CO$_2$, N$_2$, and O$_2$. The reflected part of the spectrum is accounted for by Rayleigh scattering by H$_2$, CH$_4$, N$_2$, H$_2$O, CO, CO$_2$, and O$_2$, with the corresponding refractive indices (for the scattering cross-sections) given by \citet[][]{Weber:2003}. All models are cloud-free. Cloudy models will be addressed in future work.

The mean molecular mass and heat capacity are also pre-computed for each composition on the 1060-layer $(p,T)$ grid before being used in the radiative-convective calculation. The mean heat capacities are approximated using available data for H$_2$O, CO$_2$, N$_2$, CO, CH$_4$, O$_2$, H$_2$, and performing a mass weighted average assuming these molecules are the only constituents. 

 The parameter space used is similar to the one used by \citet{Miller-Ricci:2009}. By contrast, our atmospheric composition is computed self-consistently from the equilibrium chemistry of vapor in contact with a magma ocean, rather than assuming preset abundances for the entire atmosphere. The opacities for all species are used, to the extent that they are available, resulting in a fully non-grey atmosphere. We also explore the effects of disequilibrium chemistry and photochemistry, which have not been taken into account by the study of \citet{Miller-Ricci:2009}. This work for the first time quantifies the connection between the surface and effective temperatures, thus allowing estimates of the cooling time of post-giant-impact planets. This will enable further understanding of the early surface temperature evolution of the Earth, in the context of a non-grey atmosphere. 

Atmospheric composition as a function of altitude is shown for a few representative models in Figures \ref{abunds_cc} and \ref{abunds_bse}. The influence of each molecule shown is accounted for in both the radiative transfer calculation and the underlying chemistry. Emergent high-resolution spectra are computed using the line-by-line radiative transfer code DISORT \citep{Stamnes:1988}, employing a fine opacity grid at every point in the atmosphere. DISORT can take into account an external illumination source, the solar spectrum in this case, and a Lambertian lower boundary. The effects of vertical mixing and photochemistry on the atmospheric structure are discussed in Section~\ref{diseq}.

\section{RESULTS\label{results}}

\subsection{Pressure-temperature profiles\label{sec:pt}}

\begin {figure}[h]
\hspace{-0.1\linewidth}
\includegraphics*[scale=0.7,angle=0]{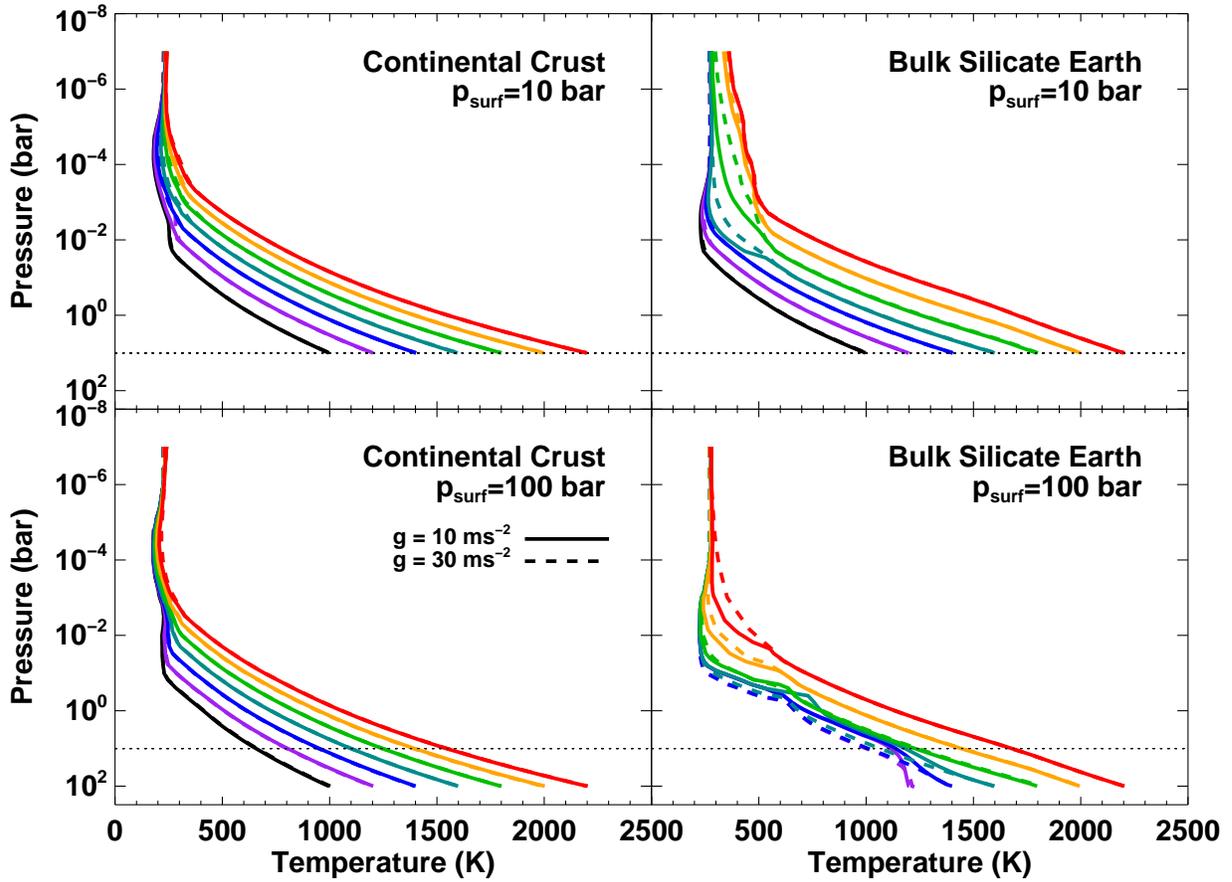}
\caption{Pressure-temperature profiles for all models, color-coded from red to black from the highest to the lowest surface temperature. The models for the Continental Crust composition are shown on the left, and those for the Bulk Silicate Earth on the right. The  models with $p_{\rm surf}=10$~bar and $p_{\rm surf}=100$~bar are shown in the upper and lower panels, respectively. A dotted line is plotted at $p_{\rm surf}=10$~bar in all panels to guide the eye. Models with $g=30$~m~s$^{-2}$ are shown with a dashed line. The effective temperatures for all models are listed in Table~\ref{tab:teff}. \label{pt_profile}}
\end{figure}

Pressure-temperature profiles for all 56 models are shown in Figure~\ref{pt_profile}. The total power radiated by the planet, or the wavelength-integrated flux, can be expressed as $\sigma_BT_{\rm eff}^4$, where $T_{\rm eff}$ is the temperature of a blackbody with the same emitted power. These effective temperatures are calculated by the atmosperic structure code and are listed in Table~\ref{tab:teff}. Because of the greenhouse effect the bulk silicate earth case with $p_{\rm surf} = 100$~bar (lower right panel in Figure~\ref{pt_profile}) cannot reach a surface temperature as cool as 1000~K (black line in other panels) at the given external radiation flux. Here, the models with $T_{\rm surf} \leq 1600$~K tend to a similar effective temperature, independent of $T_{\rm surf}$, as the atmosphere becomes optically thick. Therefore, $T_{\rm eff}$ is completely governed by the incident radiation at the top of the atmosphere, as even without any internal heating sources the surface temperature of the 100~bar atmosphere cannot drop below $\sim1200$~K. The incoming stellar radiation gets trapped, warming the surface up to the point where the peak of the blackbody function shifts to short enough wavelengths that the radiation is able to escape through existing opacity windows. The lowest surface temperature in this case is achieved through the formation of an isothermal layer at the bottom of the atmosphere, since the atmospheric composition does not allow for a steeper adiabat. Lower surface temperatures can only be reached for a lower incident flux, for example at larger star-planet separations.

\begin {figure}[h]
\hspace{-0.1\linewidth}
\includegraphics*[scale=0.7,angle=0]{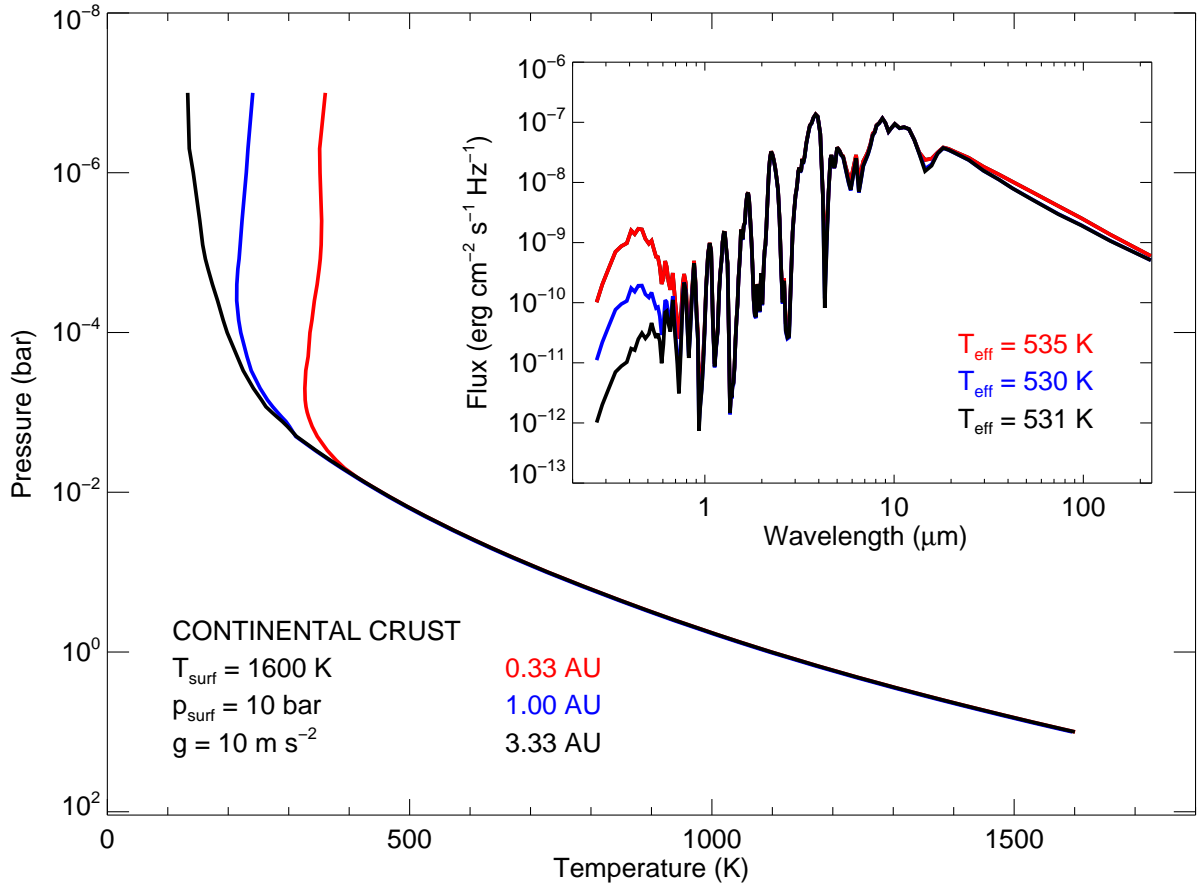}
\caption{Variation of the atmospheric pressure-temperature profile with planet-star distance. For this the Continental Crust model with $p_{\rm surf}=10$~bar, $T_{\rm surf}$=1600~K, and $g=10$~m~s$^{-2}$ was considered. The star-planet distance controls the temperature at the top of the atmosphere, with the coldest corresponding to the largest distance. The reflected light portion of the spectrum (see insert) will also be weakest for the largest distance, due to the decrease in the received flux. The effective temperature only varies by about 5~K between these models. \label{sun_dist}}
\end{figure}

 The $T-p$ profiles tend to become isothermal at the top of the atmosphere, where the temperature is controlled by the incident solar flux. This is more clearly shown in Figure~\ref{sun_dist}. In this example, using the continental crust composition with $p_{\rm surf} = 10$~bar and $T_{\rm surf} = 1600$~K, models with the same surface temperature show an increase in outer layer (top of the atmosphere) temperature with decreasing star-planet distance, while the reflected part of the spectrum is more prominent at smaller separations. For the models in Figure~\ref{sun_dist} the effective temperature $T_{\rm eff}$ remains almost constant, being determined mainly by internal heating ($T_{\rm surf}$) and atmospheric opacity, which control the IR part of the spectrum.

The effective temperatures in Table~\ref{tab:teff} are also plotted in Figure~\ref{teff}. The dependence of $T_{\rm eff}$ on $T_{\rm surf}$ has implications for the evolution and cooling of the planet after the giant impact, which will be explored further in Section~\ref{sec:evol}. Higher surface gravity models (dashed lines) have higher $T_{\rm eff}$ for any given $T_{\rm surf}$.  This is the combined result of a decreased scale height, a different placement of convection zones, and a change in the column densities of the most important absorbers. In this case, the place where the atmosphere becomes optically thin will occur deeper in the atmosphere, at a higher pressure and temperature. Due to the blanketing effect of the atmosphere, the effective temperatures are lower than the surface temperatures by factors of a few. As is often the case in planetary atmospheres the surface, equilibrium, effective, and wavelength-dependent brightness temperatures are not equal and care must be taken when interpreting any relevant temperature.  In Section~\ref{sec:mags} we compare these models to brown dwarfs of similar effective temperatures, and directly-imaged exoplanets. 

The flattening of the effective temperature curves at low $T_{\rm eff}$ seen in Figure~\ref{teff} is due to the atmosphere becoming optically thick, completely obscuring and shielding the surface, as the curves converge to the limit imposed by trapping of the incident flux from the star. This is the runaway greenhouse limit, in which case the surface temperature is completely decoupled from the observed effective temperature. Even without any internal energy sources, the stellar radiation trapped by the atmosphere determines the limiting $T_{\rm eff}\approx 265$~K, and the surface temperature cannot cool much below $\sim$1200~K, for the BSE case. The radiation limit corresponding to this $T_{\rm eff}$ is 280~W~m$^{-2}$, in line with other independent model predictions for the Earth's atmosphere \citep{Kopparapu:2013,Goldblatt:2012}.

\subsection{Atmospheric composition\label{sec:comp}}

The relative abundances of all molecular and atomic species can be computed at any point in the atmosphere, interpolating the initial equilibrium chemistry grid. Representative abundance profiles are shown in the left panels of Figures~\ref{abunds_cc} and \ref{abunds_bse} for the continental crust and bulk silicate earth cases, respectively, with $T_{\rm surf} = 1600$~K, $p_{\rm surf} = 100$~bar, and $g = 10$~m~s$^{-2}$. The right panels of these figures show the effects of vertical mixing and photochemistry, which we discuss in more detail in Section~\ref{diseq}. The pressure-temperature profiles for these models are shown by the dashed lines in the same figures, with the temperature axis at the top. For the continental crust case, the atmosphere is dominated by H$_2$O and CO$_2$, the latter  becoming most abundant above the water condensation zone. H$_2$ and CH$_4$ are secondary constituents in the upper atmosphere, while in the lower atmosphere they are replaced by HF, HCl, and SO$_2$. For the bulk silicate earth composition in equilibrium with the hot magma, H$_2$O and CO$_2$ are dominant at higher pressures, while the upper atmosphere is composed mostly of H$_2$, CH$_4$, and NH$_3$. While still important, HCl, HF, and SO$_2$ are less prominent than in the continental crust case, while NH$_3$ becomes an important opacity source. The NH$_3$ and CH$_4$ curves are shown dashed, because the formation of these species becomes quenched at higher altitude, as explained in Section~\ref{diseq}. The emerging overall picture is that in spite of a relatively cool, water-dominated spectrum, these worlds present extreme, inhospitable surface conditions, even after taking chemical disequilibrium into account.

\subsection{Emergent spectra\label{sec:spec}}

\begin {figure}[h]
\hspace{-0.08\linewidth}
\includegraphics*[scale=0.7,angle=0]{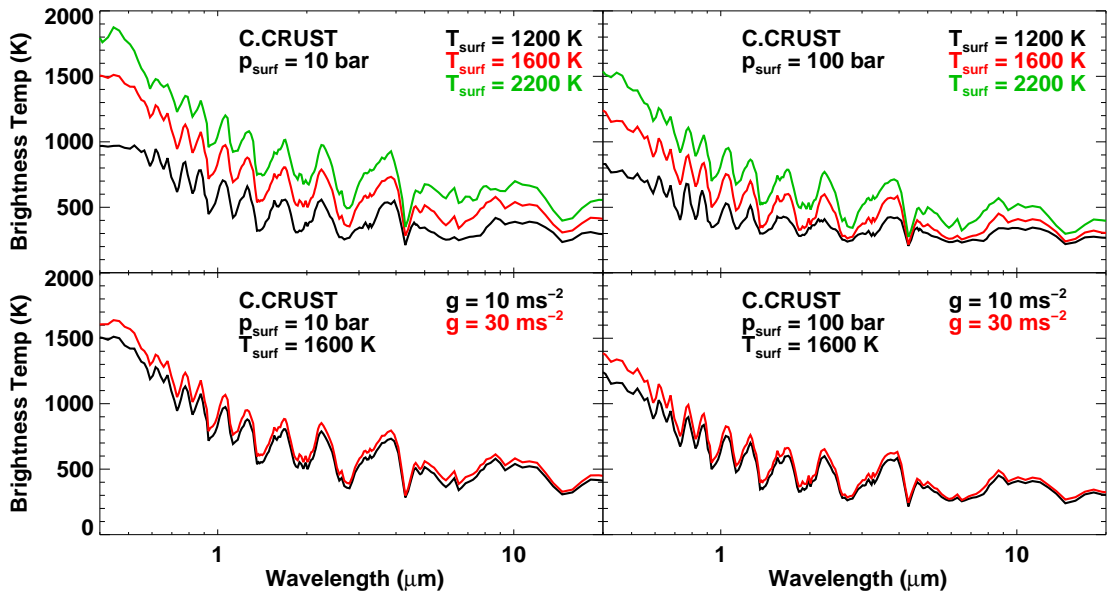}
\caption{Overview of the computed spectra for the Continental Crust models. The top panels show the comparison between models with different surface temperatures for $p_{\rm surf}=10$~bar (left) and $p_{\rm surf}=100$~bar (right). The brightness temperature scale emphasizes the drop in the outgoing radiation relative to $T_{\rm surf}$. The bottom panels compare the models with different values of $g$ at the same $T_{\rm surf}$, 1600~K in this example, for $p_{\rm surf}=10$~bar (left) and $p_{\rm surf}=100\,\rm bar$ (right). The $g=30\,\rm m\,s^{-2}$ models are slightly brighter than their $g=10$~m~s$^{-2}$ equivalents. For the individual labeling of major features in these spectra see Figure~\ref{cc_comp}. \label{compare_cc}}
\end{figure}

\begin {figure}[h]
\hspace{-0.08\linewidth}
\includegraphics*[scale=0.7,angle=0]{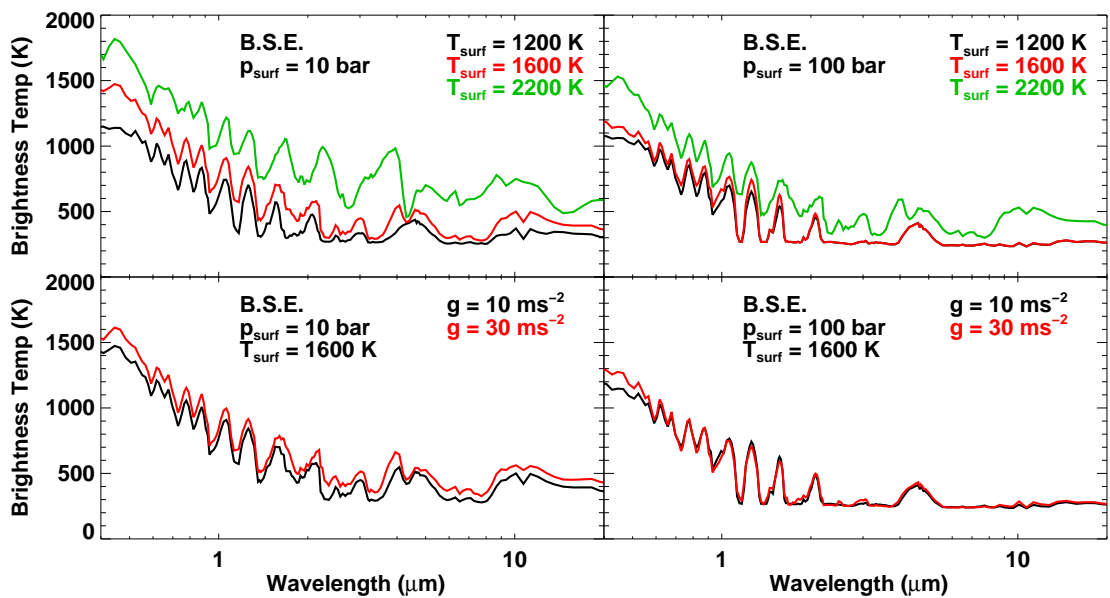}
\caption{Same as Figure~\ref{compare_cc} for the Bulk Silicate Earth case. The differences between the $g=10$ and $g=30$~m~s$^{-2}$ models are minimal. For the individual labeling of major features in these spectra see Figure~\ref{bse_comp}. \label{compare_bse}}
\end{figure}

 An overview of the emergent low-resolution spectra is given in Figures~\ref{compare_cc} and \ref{compare_bse}, for the continental crust and the bulk silicate earth compositions, respectively. The upper panels compare spectra from models with the same surface pressure and gravity, but different surface temperatures, while the lower panels compare the spectra for models with the same surface temperature and pressure, but different surface gravities. As discussed above, models with higher $g$ have a slightly higher brightness temperature. The brightness temperature scale highlights the fact that the hot surface is obscured at virtually all wavelengths in these models. Even the brightest features in the near- to mid- infrared region have a brightness temperature a factor of $\sim$2 lower than $T_{\rm surf}$ for the continental crust case, which becomes a factor of $\sim$3 for the bulk silicate earth models. The spectral features change little with surface temperature for the continental crust composition (Figure~\ref{compare_cc}), but are progressively altered for the BSE models, as the composition changes with $T_{\rm surf}$. For the BSE case with $p_{\rm surf}=100$~bar (upper right panel in Figure~\ref{compare_bse}), models with $T_{\rm surf} \leq 1600$~K lead to virtually indistinguishable spectra. This is consistent with the previous conclusions about the effective temperature, and reflects the fact that the atmosphere has the same structure down to the level where the radiation is escaping from.

\begin {figure}[h]
\hspace{-0.1\linewidth}
\includegraphics*[scale=0.7,angle=0]{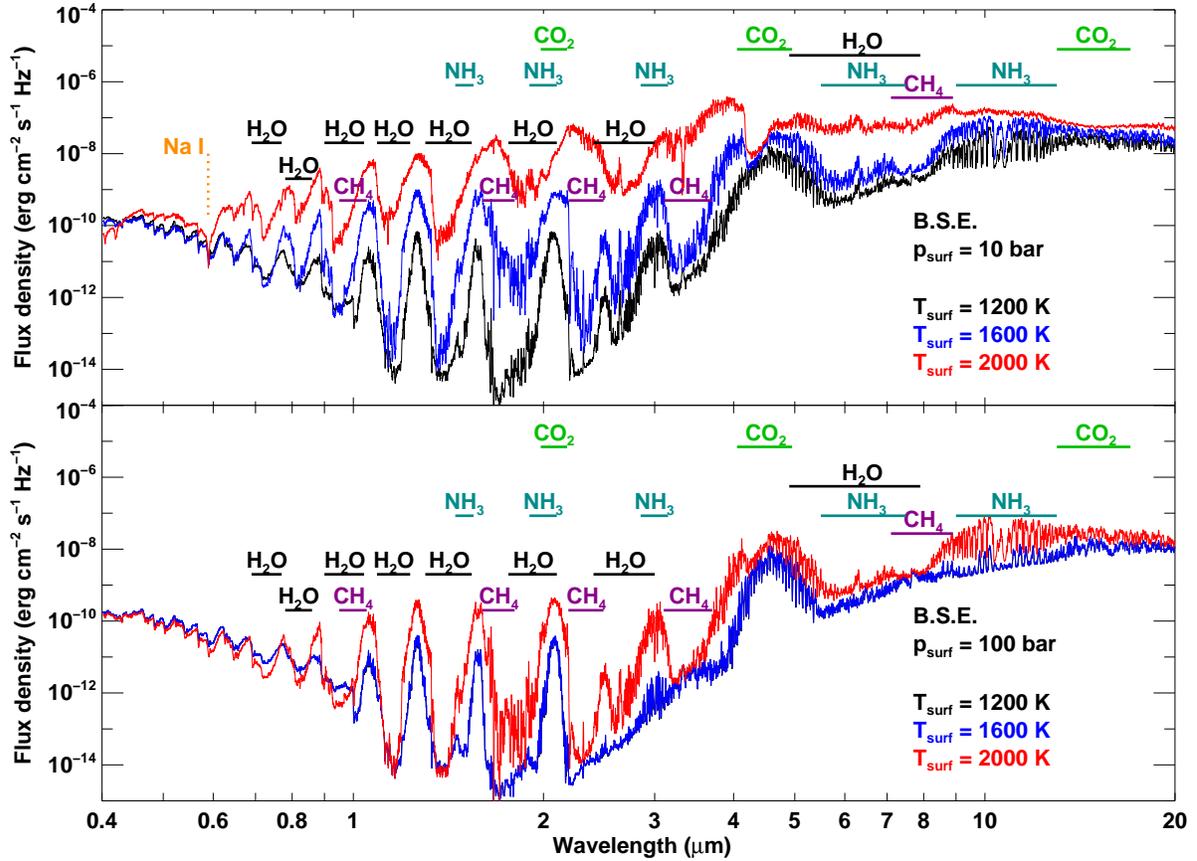}
\caption{Comparison between the high-resolution spectra for the Continental Crust models as a function of surface temperature, for $p_{\rm surf}=10$~bar (top panel) and $p_{\rm surf}=100$~bar (bottom panel). In all cases, the dominant features remain comparatively unchanged, with the amount of CH$_{4}$ increasing slowly with decreasing $T_{\rm surf}$, and the CO$_{2}$ contribution decreasing relative to water. \label{cc_comp}}
\end{figure}

\begin {figure}[h]
\hspace{-0.1\linewidth}
\includegraphics*[scale=0.7,angle=0]{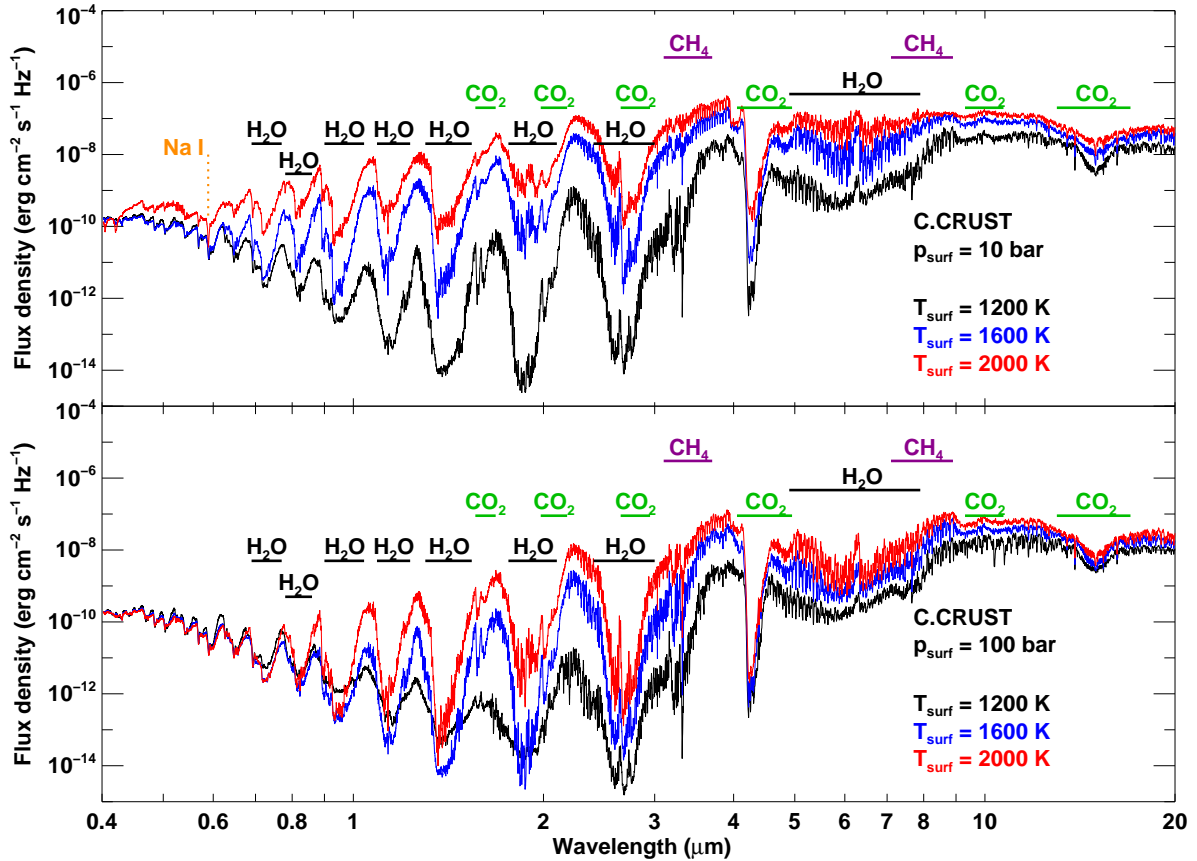}
\caption{Same as Figure~\ref{cc_comp}, for the Bulk Silicate Earth case. For $p_{\rm surf}=10$~bar (top panel), as $T_{\rm surf}$ decreases, the \ion{Na}{1} D lines and the CO$_{2}$ features become less prominent, while the CH$_{4}$ and NH$_{3}$ contributions increase. For $p_{\rm surf}=100$~bar (top panel), the CO$_{2}$ is only apparent at large $T_{\rm surf}$, while NH$_{3}$ becomes one of the dominant sources of opacity for lower $T_{\rm surf}$. Note that for $p_{\rm surf}=100$~bar and $T_{\rm surf}\leq 1600$~K, the atmosphere becomes optically thick, and the models cannot be further distinguished.  \label{bse_comp}}
\end{figure}

High-resolution spectra are obtained using the line-by-line radiative transfer code DISORT \citep{Stamnes:1988}, with finely sampled opacities and equilibrium abundances at every atmospheric layer as inputs. Figures~\ref{cc_comp} and \ref{bse_comp} show high-resolution spectra for the continental crust and bulk silicate earth case, respectively, at $R \sim 500-600$ resolution. The two panels compare the spectra at constant surface pressure and gravity, for increasing surface temperatures. The most important features are identified. As expected from the atmospheric composition, in the CC case the spectrum is dominated by H$_2$O and CO$_2$. As the surface temperature decreases, CH$_4$ features become more prominent, while the CO$_2$ features weaken. The \ion{Na}{1} $D$ lines at 0.59~$\mu$m are apparent at highest $T_{\rm surf}$ and low $p_{\rm surf}$ (upper panel). The spectra of BSE models (see Figure~\ref{bse_comp}), are also dominated by water absorption but, in contrast to the CC case, the CO$_2$ contribution is much diminished, being only significant for the higher $T_{\rm surf}$ models, while CH$_4$ becomes more important, comparable to water at lower temperatures for $p_{\rm surf} = 10$~bar, and especially predominant in all models for $p_{\rm surf} = 100$~bar. Interestingly, under the assumption of equilibrium chemistry, NH$_3$ is a significant absorber and an abundant molecule for the BSE case, with lines that become completely optically thick for most of the $p_{\rm surf}=100$~bar models, and for low $T_{\rm surf}$ in general. The combination of H$_2$O, CH$_4$, and NH$_3$ lines covering most of the infrared region makes these planets heavily obscured and subject to a strong greenhouse effect. For emphasis, in Figure~\ref{cc_bse_comp} we compare directly the BSE and CC spectra shown in Figures~\ref{bse_comp} and \ref{cc_comp}. This figure clearly displays the above mentioned differences between the continental crust and the bulk silicate earth models. It also emphasizes the similarity of the two compositions for $p_{\rm surf}=10$~bar, in which case the equilibrium chemistry for BSE does not favor significant production of CH$_4$ and NH$_3$, while the presence of larger amounts of CO$_2$ and SO$_2$ characterizes the CC spectrum. 

\begin {figure}[h]
\begin{minipage}{\linewidth}
      \centering
\begin{minipage}{0.34\linewidth}
\hspace{-0.1\linewidth}
\includegraphics*[width=1.5\linewidth]{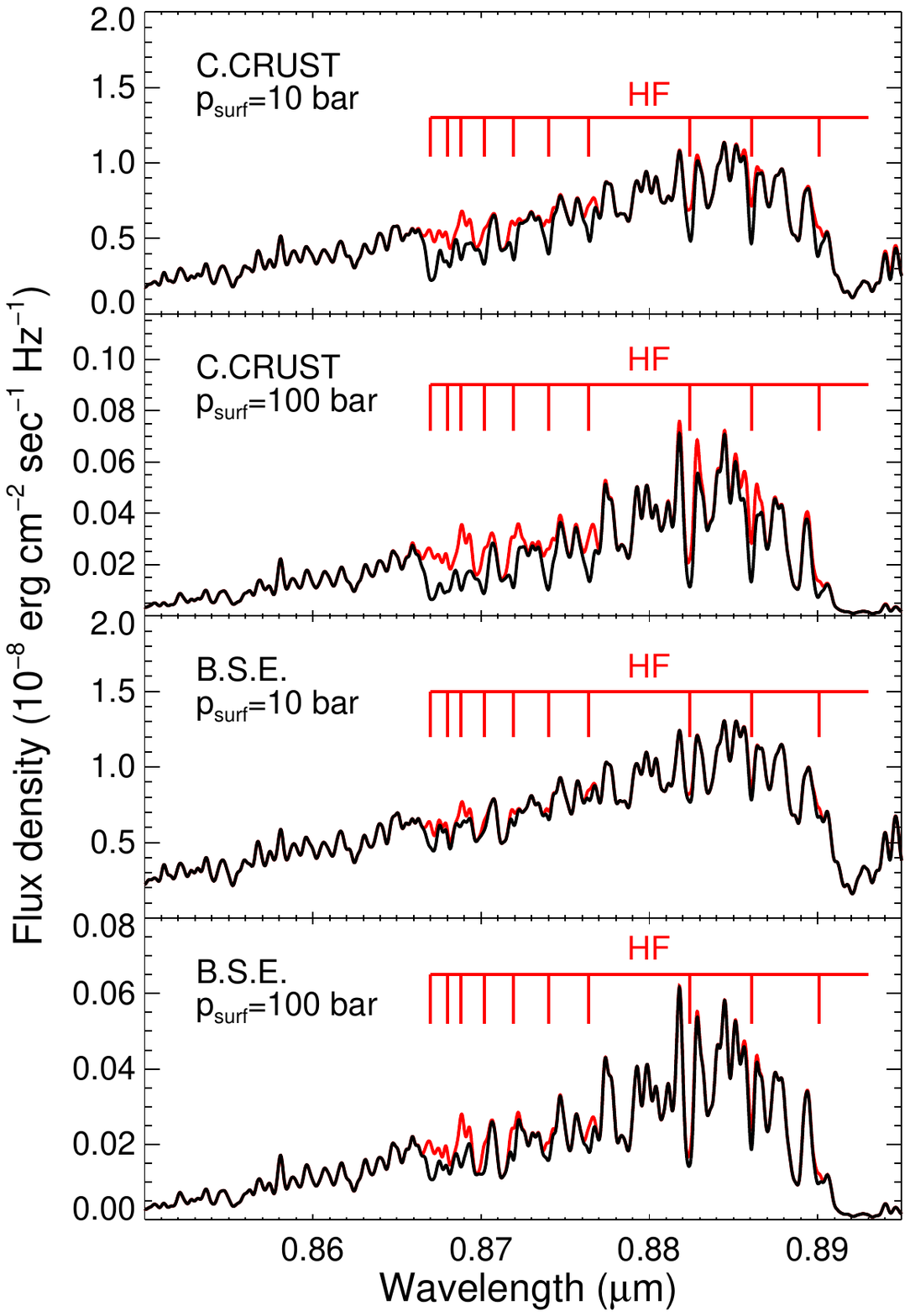}
\end{minipage}
      \hspace{-0.05\linewidth}
      \begin{minipage}{0.34\linewidth}
\includegraphics*[width=1.5\linewidth]{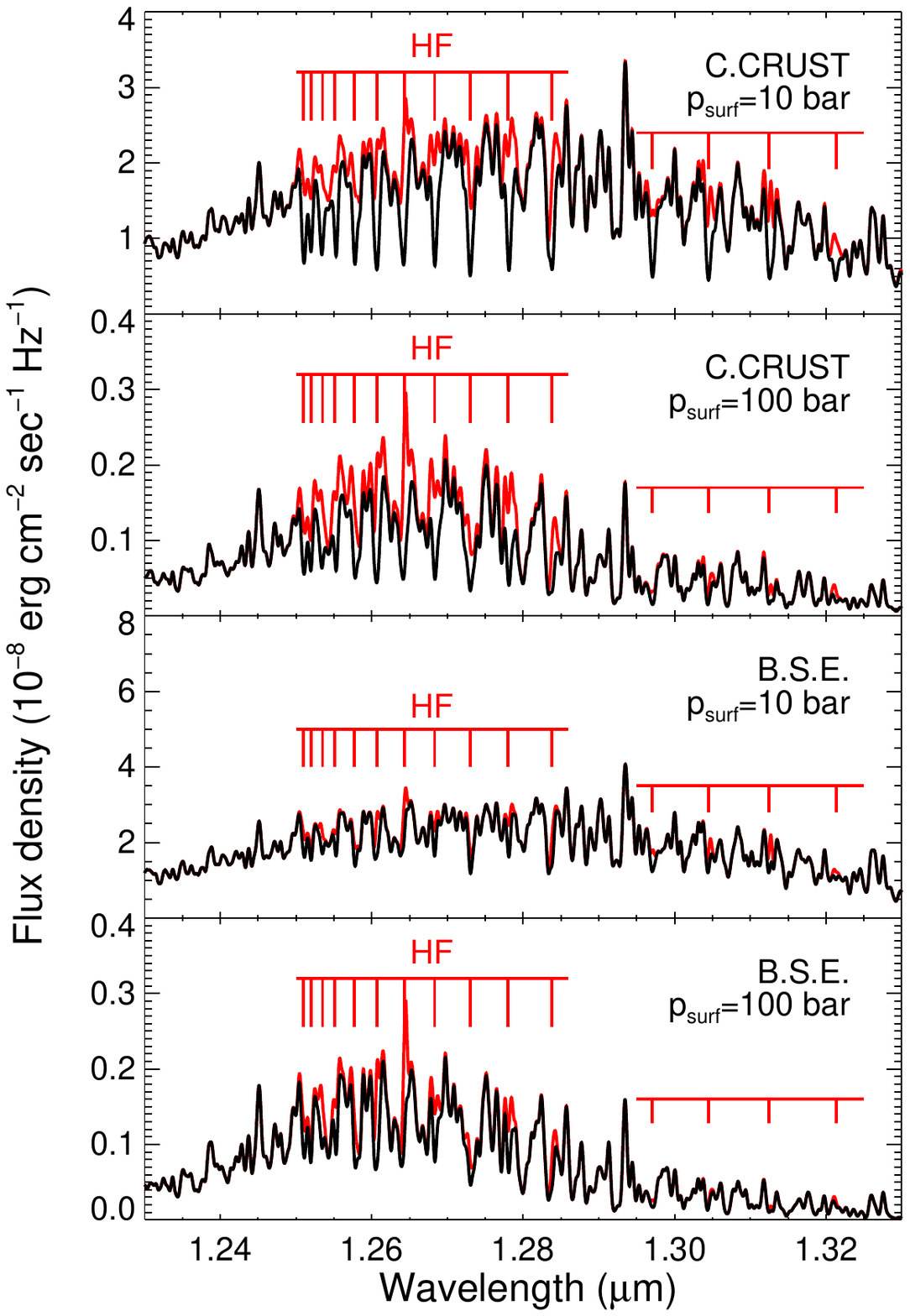}
\end{minipage}
      \hspace{-0.03\linewidth}
      \begin{minipage}{0.34\linewidth}
\includegraphics*[width=1.5\linewidth]{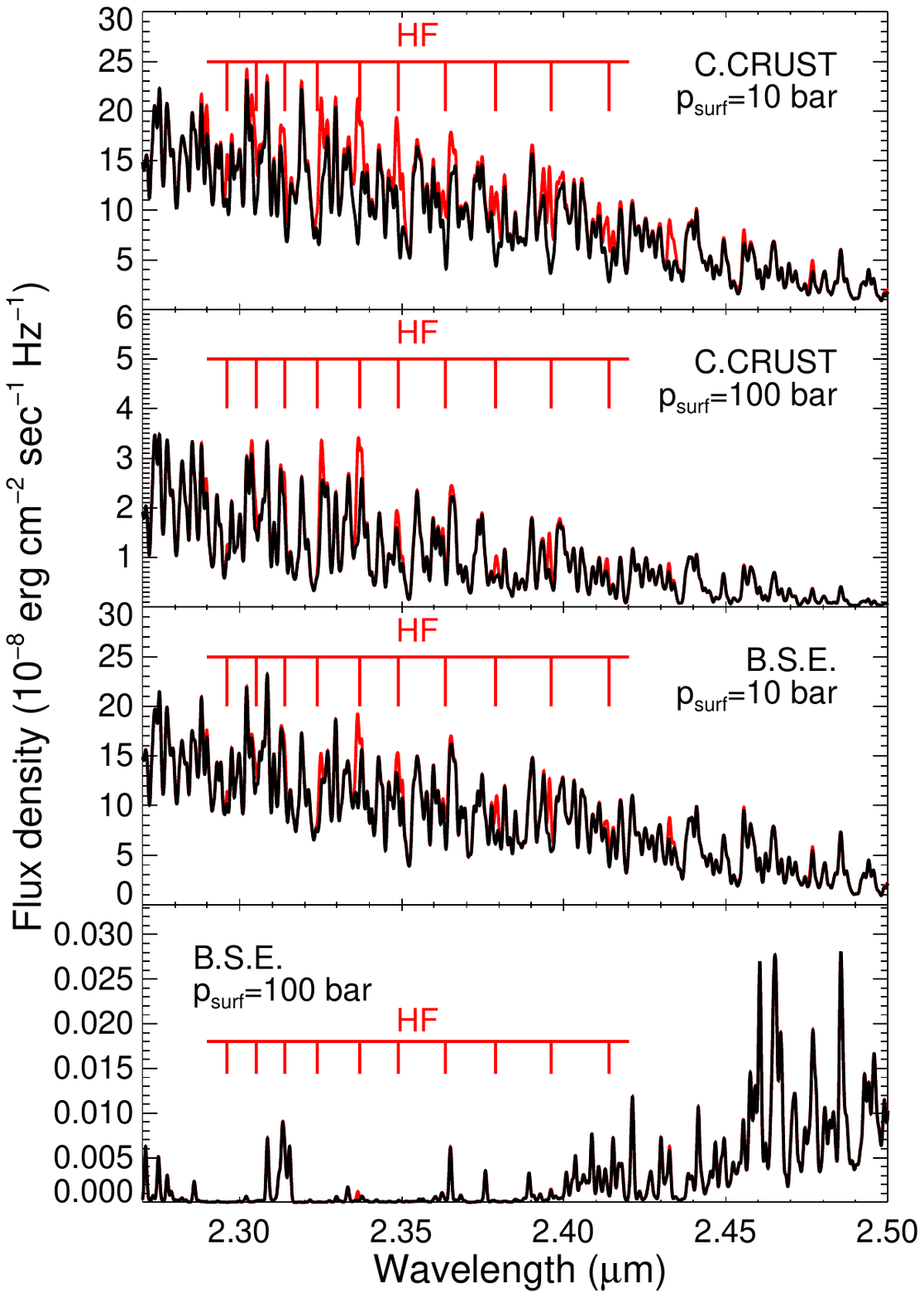}
\end{minipage}
  \end{minipage}
\caption{HF features (red) in the spectra of post-giant impact planets, for models with a surface temperature of 2200~K and $g=10$~m~s$^{-2}$. The black line shows the original high-resolution spectrum, while the red lines show the resulting high-resolution spectrum after removing the HF opacity from the mix. The HF presence is more prominent in the Continental Crust case. \label{hf}}
\end{figure}

\begin {figure}[h]
\begin{minipage}{\linewidth}
      \centering
\begin{minipage}{0.34\linewidth}
\hspace{-0.1\linewidth}
\includegraphics*[width=1.5\linewidth]{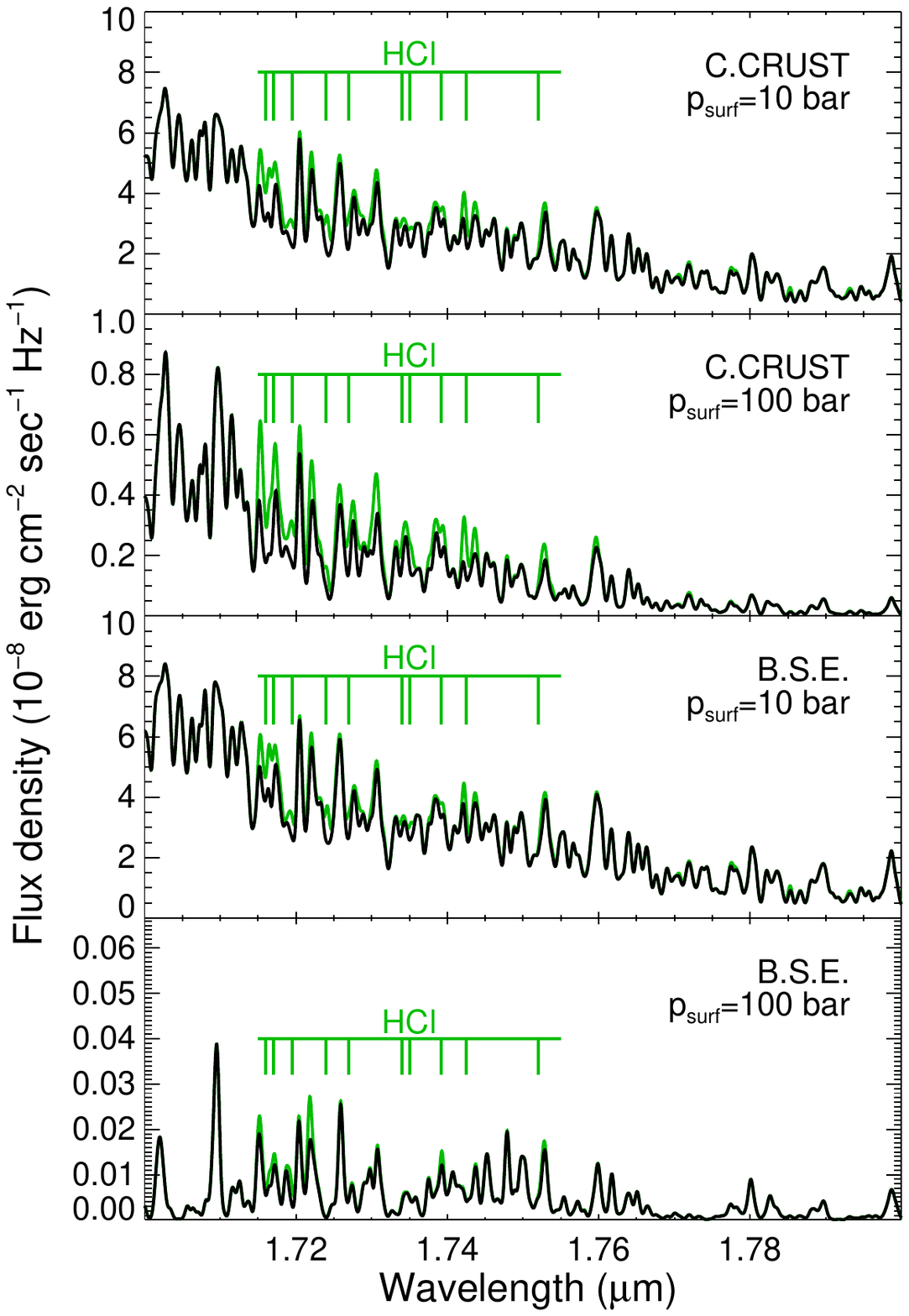}
\end{minipage}
      \hspace{-0.05\linewidth}
      \begin{minipage}{0.34\linewidth}
\includegraphics*[width=1.5\linewidth]{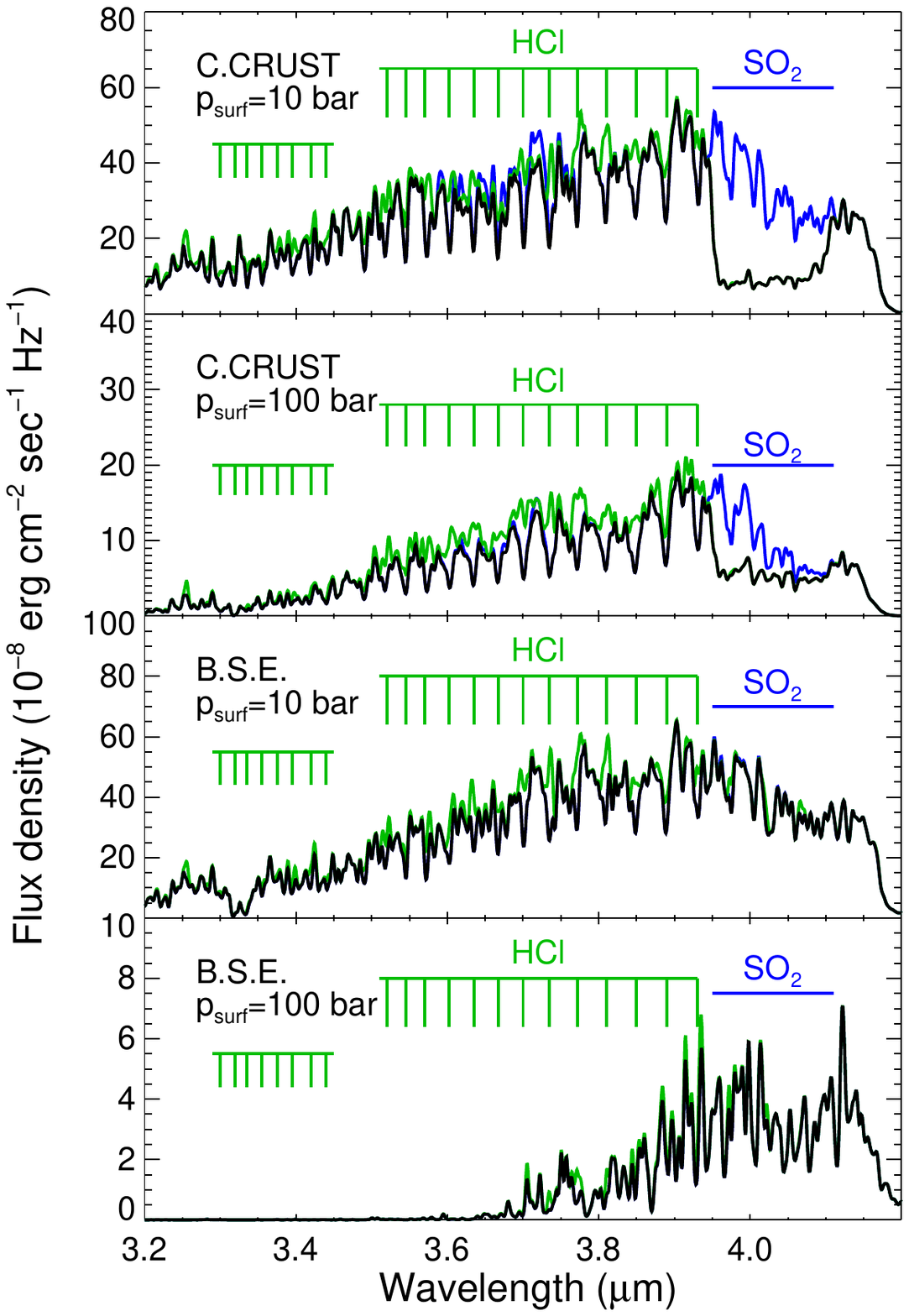}
\end{minipage}
      \hspace{-0.03\linewidth}
      \begin{minipage}{0.34\linewidth}
\includegraphics*[width=1.5\linewidth]{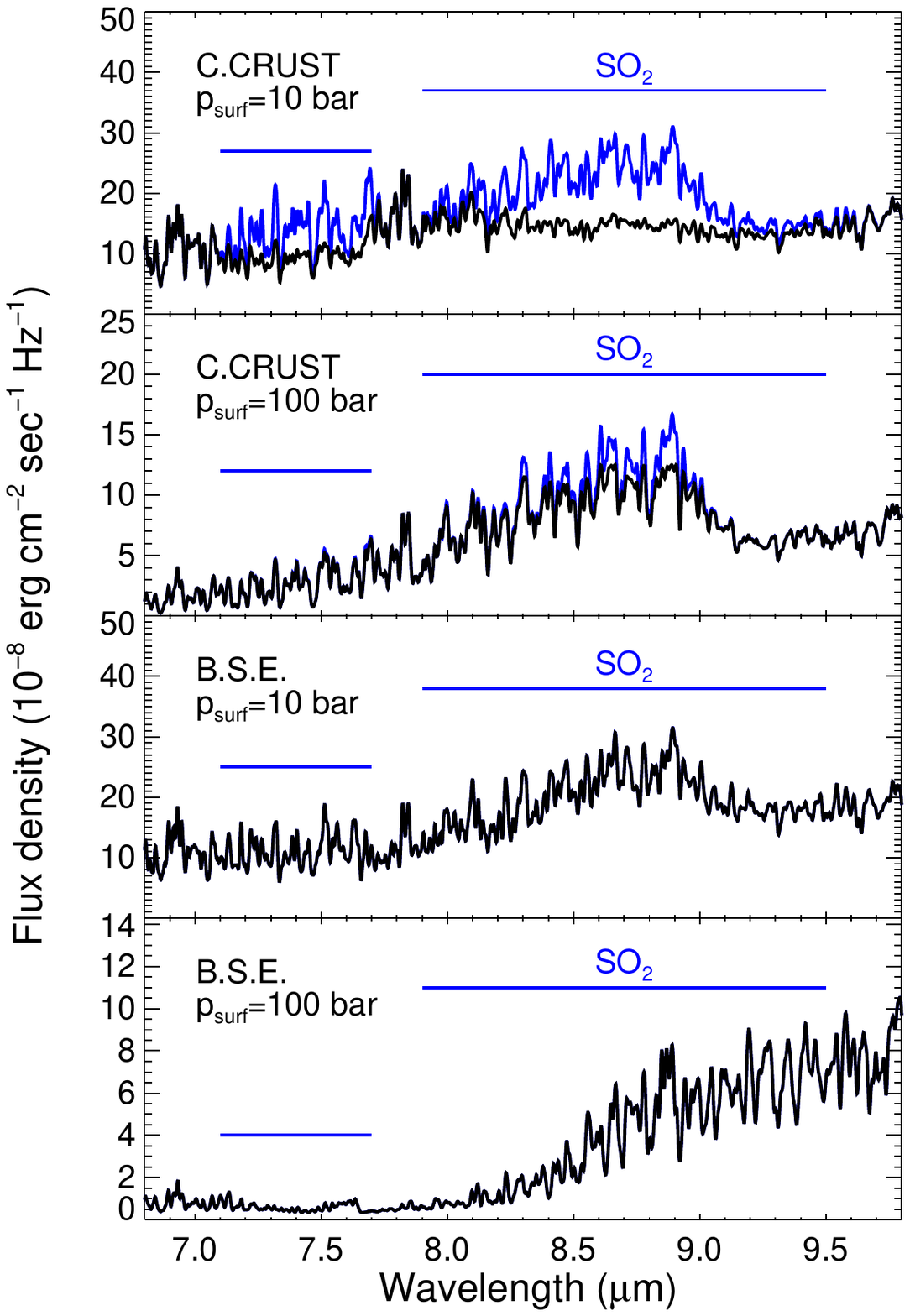}
\end{minipage}
  \end{minipage}
\caption{HCl (green) and SO$_{2}$ (blue) features in the spectra of post-giant impact planets, for models with a surface temperature of 2200~K and $g=10$~ms$^{-2}$. As in Figure~\ref{hf}, the black line shows the original high-resolution spectrum, while the green and blue lines show the resulting high-resolution spectra after removing the HCl and SO$_{2}$ opacities, respectively. The presence of SO$_{2}$ features is characteristic to the Continental Crust models. \label{hclso2}}
\end{figure}

The lower atmosphere constituents mentioned in Section~\ref{sec:comp} also have discernable features in the high-resolution spectra, as some strong transitions fall in the window regions. The HF signature is shown in red in Figure~\ref{hf}, while the HCl and SO$_2$ features are shown in green and blue, respectively, in Figure~\ref{hclso2}. As these constituents become prominent only at high temperatures, we show only the spectra for $T_{\rm surf} = 2200$~K. The spectra shown in color are computed after the opacity of the corresponding molecule has been removed from the mix. This helps guide the eye to the relative importance of the HF, HCl, and SO$_2$ features, given the abundance of overlapping lines present in these spectra. For HF we distinguish a band edge at 0.867~$\mu$m, and a strong line pattern between 1.25$-$1.29~$\mu$m. HCl shows a similar pattern between $3.5-3.9$~$\mu$m, while SO$_2$ has broad-band signatures, most notably one starting at 3.95~$\mu$m, near the wing of the CO$_2$ band (for context, see Figure~\ref{cc_bse_comp}). From Figure~\ref{hclso2} it is apparent that SO$_2$ can be considered as a marker for the continental crust composition, and could be used to distinguish between a post-giant-impact planet and a cool water world. The HF and HCl lines are also weaker in the bulk silicate earth models.

We note that HCl and HF have been detected in the near-infrared opacity windows at Venus \citep{bezard:1990}. These molecules appear to be robust indicators of a hot, greenhouse atmosphere and could help distinguish such atmospheres from $\rm CO_2$-dominated atmospheres with cooler surface temperatures. The chemistry models of \citep{Schaefer:2011} predict the presence of important amounts of HCl and HF in the atmospheres of hot rocky planets, and show that a cooler surface leads to less HCl. By calculating the $T-p$ profiles, we find that these molecules are also potentially detectable in exoplanetary atmospheres.
 
\begin {figure}[h]
\centering
\includegraphics*[scale=0.6,angle=0]{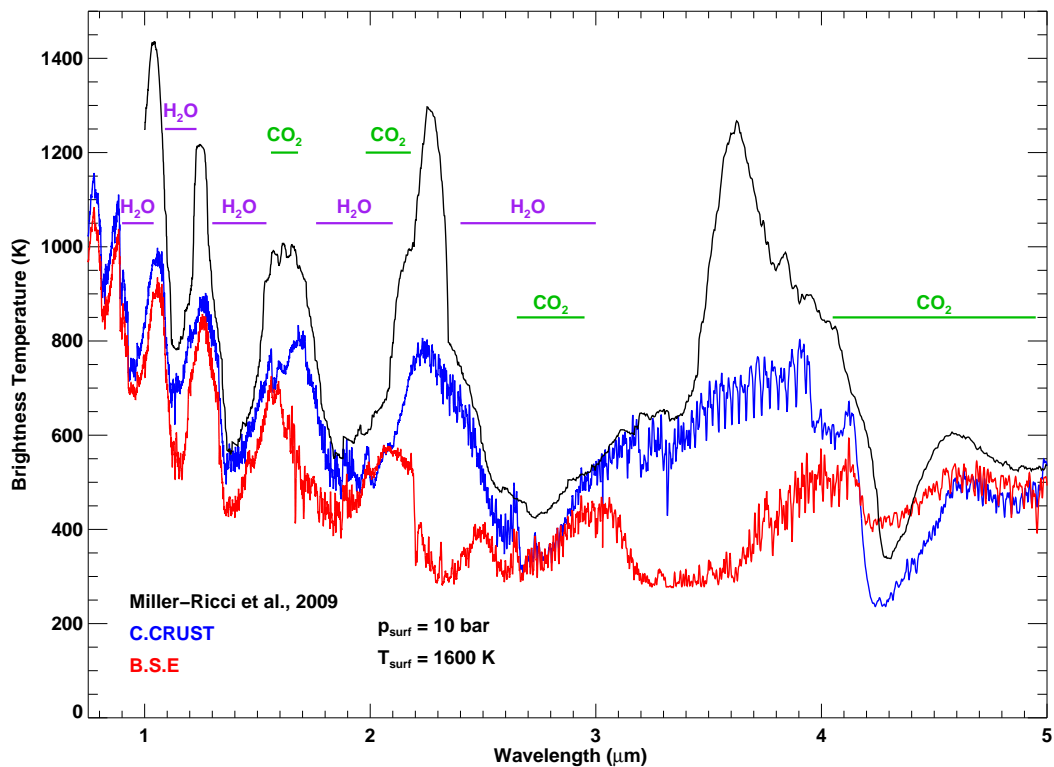}
\caption{Comparison between the Continental Crust (blue), the Bulk Silicate Earth (red), and a previous model spectrum from \citet{Miller-Ricci:2009}, composed of 90\% H$_2$O and 10\% CO$_2$. All models have roughly the same surface pressure and temperature. Due to additional opacity sources, the current models predict substantially fainter emission than the ones described in \citet{Miller-Ricci:2009}. \label{compare_eliza}}
\end{figure}

 Finally, we compare the current models to the previous result of \citet{Miller-Ricci:2009} in Figure~\ref{compare_eliza}. All models in this figure have $p_{\rm surf} = 10$~bar, $T_{\rm surf} = 1600$~K, and $g = 10$~m~s$^{-2}$. The \citet{Miller-Ricci:2009} models contains 90\% H$_2$O and 10\% CO$_2$, and the predicted spectrum (black) is more similar to the continental crust model (blue), which does not contain the large CH$_4$ absorption features present in the bulk silicate earth case (red). However, due to the additional opacity sources considered, the spectral peaks predicted by the current models have brightness temperatures 1/3 cooler on average than the previous estimates. The consequences for direct detection are discussed in Section~\ref{sec:detect}. 
 
\subsection{Model Photometry\label{sec:mags}}

For a subset of the models, simulated infrared photometry was calculated using the model emergent spectra, for the abundances predicted by chemical equilibrium. Absolute magnitudes were calculated in the near-infrared in \emph{Y}, \emph{J}, \emph{H}, and \emph{K} bands and in the mid-infrared in \emph{L$^{\prime}$}, \emph{M$^{\prime}$}, \emph{Spitzer} IRAC bands 1--4, and \emph{WISE} \emph{W1}, \emph{W2}, \emph{W3}, and \emph{W4}. The radii were taken to be 1 \re\ ($g$=10 m s$^{-2}$) and 1.75 \re\ ($g$=30 m s$^{-2}$), as expected for an Earth-like composition (Eric Lopez, personal communication). The model photometry is shown in Table~\ref{magtable}. 

\begin{figure}
\centering
 \includegraphics[width=6.in]{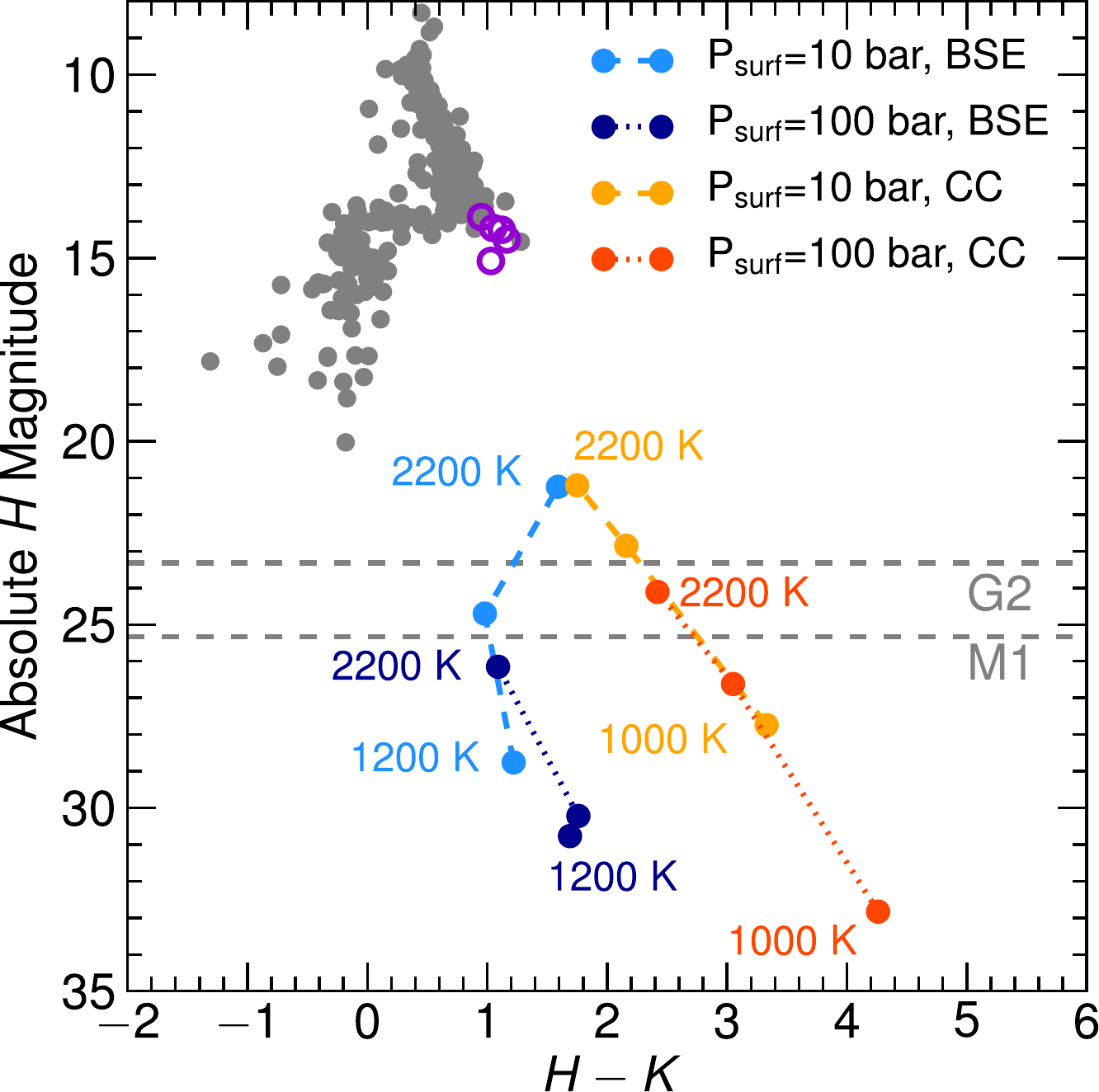}
 \caption{Color-magnitude diagram of post-impact Earths. $H-K$ color vs. absolute $H$ magnitude for a selection of models. The continental crust composition atmosphere are shown in blue and the bulk silicate Earth models are shown in red. Two surface pressures (10 and 100 bar) are shown; all models have surface gravities of 30 m s$^{-2}$. Observed brown dwarfs are shown as grey circles; directly imaged planets (4 HR 8799 planets \citep{Skemer:2012, Marois:2010} and 2M1207b \citep{Patience:2010}) are shown as violet open circles. Estimated magnitude limits for a planet around a G2 and M1 dwarf observed with a 30-m class telescope capable of 10$^{-8}$ contrast are shown as dashed grey lines; models above the line should be observable with a thirty-meter telescope. The magnitudes and colors for the post-giant impact planets are computed using the equilibrium chemistry. After correcting for vertical mixing and photochemistry, the BSE points would move closer to the CC models.}
   \label{colormag} 
\end{figure}

Figure \ref{colormag} shows $H-K$ color vs. absolute $H$ magnitude for a selection of model post-impact Earths; observed brown dwarfs and giant planets plotted for reference. The models with $p_{\rm surf}$=10~bar are significantly brighter than the $p_{\rm surf}$=100~bar models for similar surface temperatures. With the exception of the very hottest models which have similar colors, the continental crust composition atmospheres are significantly redder in infrared color $H-K$ than the bulk silicate Earth composition atmospheres. This suggests that photometry alone could potentially differentiate between the compositions studied here. 

It is also clear from Figure~\ref{colormag} that post-impact Earths are distinct in color and magnitude from brown dwarfs and giant planets; the differences are due to the larger variety of gaseous absorbers present in their atmospheres. For the hottest surface temperatures studied (2200~K), the absolute magnitude in $H$-band is roughly equal to that of the coolest brown dwarfs observed to date, but the post-impact Earths are significantly redder in infrared color. This should make it possible to identify such objects from photometry alone.

\section{Disequilibrium Processes\label{diseq}}

The abundance profiles shown in the lefthand panels of Figures~\ref{abunds_cc} and \ref{abunds_bse} have been calculated assuming complete chemical equilibrium between gases, melts, and solids, i.e., the gases in each atmospheric layer are in equilibrium with each other and also with the planetary surface. In this section we consider the effects of dis-equilibrating processes such as vertical mixing and photochemistry on the computed model atmospheres.

\subsection{Vertical Mixing\label{sec:quench}}

Sufficiently rapid vertical mixing rates relative to chemical reaction rates cause departures from chemical equilibrium in the atmospheres of gas giant planets in the solar system, extrasolar gas giant planets and brown dwarfs \citep[e.g.,][]{Prinn:1977,Fegley:1985,Fegley:1986,Fegley:1988,Fegley:1994,Fegley:1996,Lodders:2006}.
At sufficiently high temperatures and pressures in the convective regions of planetary atmospheres, chemical reactions proceed rapidly relative to vertical mixing and the chemical lifetimes ($t_{\rm chem}$) are shorter than the vertical mixing times ($t_{\rm dyn}$). In this region $t_{\rm chem}<t_{\rm dyn}$, and chemical equilibrium is attained. As temperature and pressure decrease with increasing altitude in the convective region, reaction rates decrease exponentially and eventually a critical altitude level at which $t_{\rm chem}=t_{\rm dyn}$ is reached. This critical level, which is different for each chemical reaction because of their different reaction rate constants, is the quench level at which chemical equilibrium is frozen in. At higher altitudes (and lower $T$ and $p$), chemical reactions proceed significantly slower and chemical lifetimes are significantly longer than vertical mixing times ($t_{\rm chem}>t_{\rm dyn}$). Chemical equilibrium is not attained in this disequilibrium region and the chemical equilibrium abundances established at the quench level prevail in the absence of further dis-equilibration due to photochemistry. While a complete analysis of disequilibrium chemistry for our atmospheric models is beyond the scope of this paper, we can identify possible deviations from chemical equilibrium, and their effect on the predicted spectra.

Following \citet{Gierasch:1985}, we calculate the vertical eddy diffusion coefficient $K_{zz}$ from mixing length theory as a function of height in the atmosphere (or equivalently pressure):

\begin{equation}
K_{zz}=\frac{1}{3}h_z\left(\frac{R_{gas} {\dot Q_{cz}}}{\mu_z \rho_z c_{pz}}\right)^{1/3},
\end{equation}

\noindent where $h_z$ is the scale height, $R_{gas}$ is the ideal gas constant, $\mu_z$, $\rho_z$, $c_{pz}$, are the mean molecular weight, density and specific heat capacity of the gas, ${\dot Q_{cz}}$ is the convective heat flow, and the index $z$ is used to designate that these quantities vary with height. The convective heat flow ${\dot Q_{cz}}$ is calculated as the difference between the total atmospheric energy flux ($\sigma_B T_{\rm eff}^4$) and the integrated radiative flux at any particular position in the atmosphere, as given by our converged model. The dynamical timescale is then $t_{\rm dyn}=h_z^2/K_{zz}$. We find that for the hottest models, which are most likely to be observed, the dynamical timescale for mixing from the tropopause to the surface varies between  2 and 20 hours for the BSE, and between 1 and 15 hours for the CC case. Thus, the gas will be mixed and in contact with the hot magma present at the surface on a day-like timescale.

The bulk of the atmosphere for the CC and BSE compositions consists of H$_2$O and CO$_2$, which are found throughout the atmosphere. Other species, such as CH$_4$ and NH$_3$, which play an important role especially in the BSE models, are most abundant at low temperatures, and their abundances can more easily be affected by disequilibrium chemistry. For these two species, we consider the kinetics of the reactions:

\begin{eqnarray}\label{rch4}\nonumber
& {\rm CH_{4} + OH \rightarrow CH_{3} + H_{2}O} \\
& k_1=2.6\times10^{-17}(T^{1.83})\mathrm{exp}(-1400/T), [{\rm cm}^3{\rm s}^{-1}]
\end{eqnarray}
and
\begin{eqnarray}\label{rnh3}\nonumber
& {\rm NH_{3} + OH \rightarrow NH_{2} + H_{2}O } \\
& k_2=8.31\times10^{-17}(T^{1.6})\mathrm{exp}(-480/T),
\end{eqnarray}

\noindent with the corresponding reaction rates, $k_1$ and $k_2$. The rate constants for reactions \ref{rch4} and \ref{rnh3} are from \citet{Baulch:1992,Cohen:1991}, respectively. A comparison of the t$_{chem}$ and t$_{dyn}$ values shows that both reactions quench in the $900-1000$~K range for the CC and BSE model atmospheres. As a result, the atmospheric abundances of CH$_4$ and NH$_3$ at lower temperatures are frozen in at their values at the $900-1000$~K isothermal region in the atmosphere. These abundances are negligible (Figures ~\ref{abunds_cc} and \ref{abunds_bse}), and therefore the equilibrium abundance curves will no longer be valid (dashed lines). The kinetic inhibition of CH$_4$ and NH$_3$ formation also means that the CO$_2$ and N$_2$ abundance curves are quenched at their high temperature values, i.e., the abundances of CO$_2$ and N$_2$ are not reduced by formation of significant amounts of methane and ammonia. In consequence, since N$_2$ is not an important absorber, CO$_2$ will replace CH$_4$ and NH$_3$ as the dominant opacity source after water.

\subsection{Photochemistry}

We use a new 1-D atmospheric photochemistry code that is a hybrid between a ``hot Jupiter" photochemistry code \citep{Zahnle:2009,Miller-Ricci:2012} and a conventional terrestrial code \citep{Zahnle:2006b,Claire:2006}. Vertical mixing is parameterized by the eddy diffusion coefficient 
$K_{zz}$ determined by the atmosphere structure 
model (Section~\ref{sec:quench}). The new code currently includes about 1100 reactions of 96 small species (molecules and free radicals) made from H, O, C, N, S, Na, Cl.
Missing from the model are species of F and K, the important omissions are 
likely to be HF and KCl. Every chemical reaction in the photochemistry code is balanced by the corresponding reverse reaction, with the reverse reaction rate determined by
thermodynamic equilibrium. This ensures that the kinetics model will 
relax to thermochemical equilibrium in the absence of vertical mixing
and radiation. The model includes H$_2$O, H$_2$SO$_4$, S$_8$, NaCl, NaOH, NaCN, and 
Na$_2$S aerosols. The aerosols condense, fall, and evaporate, but chemistry on grain 
surfaces is not included. Nor have we yet to fully implement dissolution of species or aerosols into cloud droplets.

\begin {figure}[h]
\hspace{-0.1\linewidth}
\includegraphics*[scale=0.7,angle=0]{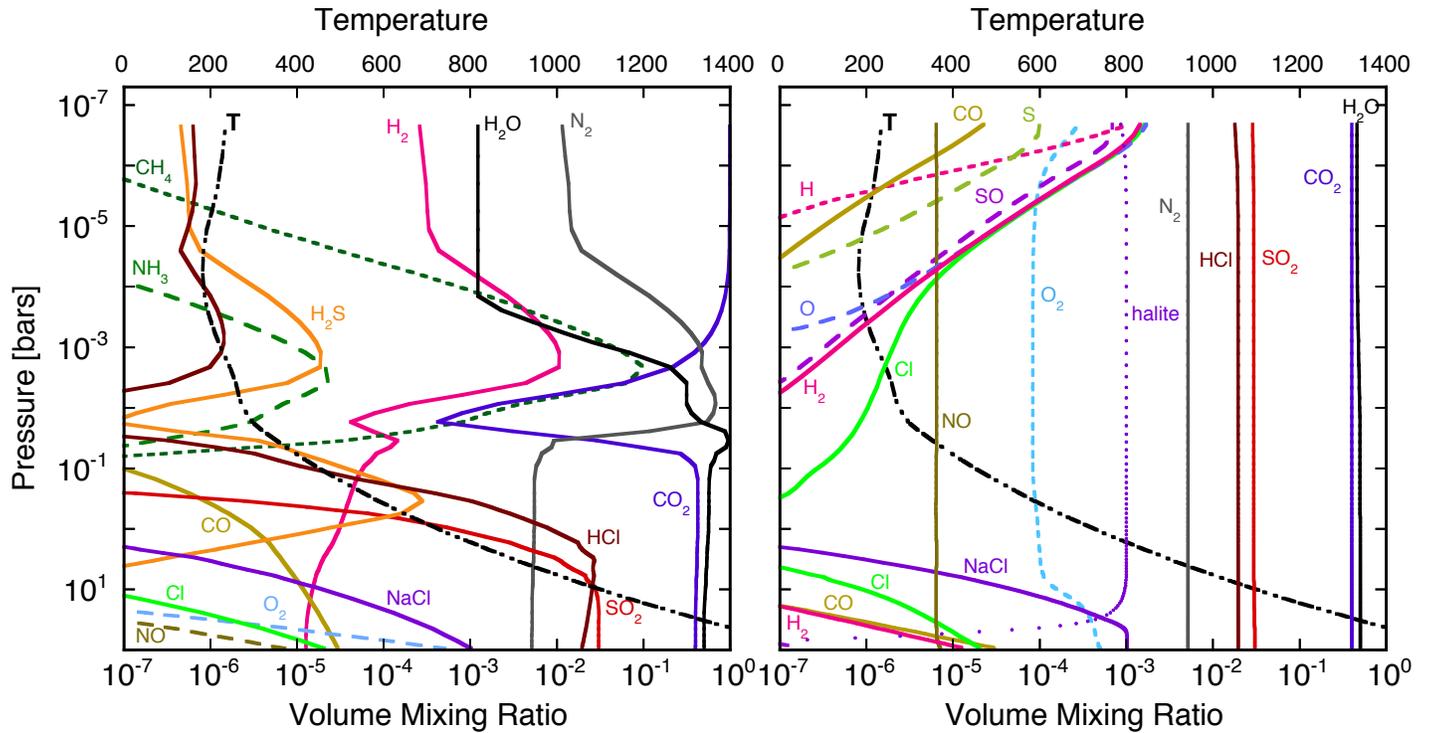}
\caption{Chemical compositions of 100~bar atmospheres over a 1600~K surface in chemical equilibrium with the continental crust composition. The pressure-temperature profile for this model is shown with a dash-dot black line, with the corresponding temperature axis at the top. Left: Each level of the atmosphere equilibrates with the surface. Composition is sensitive to temperature. Geochemical volatiles Na, Cl, and S enter minerals. The profiles for CH$_4$ and NH$_3$ are shown with dashed lines, since their abundances become quenched at temperatures between 900$-$1000~K (see Section~\ref{sec:quench}). Right: Results from the photochemical/kinetics model using $K_{zz}$ computed from the convective heat flux. Composition is dominated by mixing at middle altitudes and photolysis at high altitudes. Hydrochloric acid, salt (halite), and various sulfur species are mixed through the atmosphere. \label{abunds_cc}}

\end{figure}

\begin {figure}[h]
\hspace{-0.1\linewidth}
\includegraphics*[scale=0.7,angle=0]{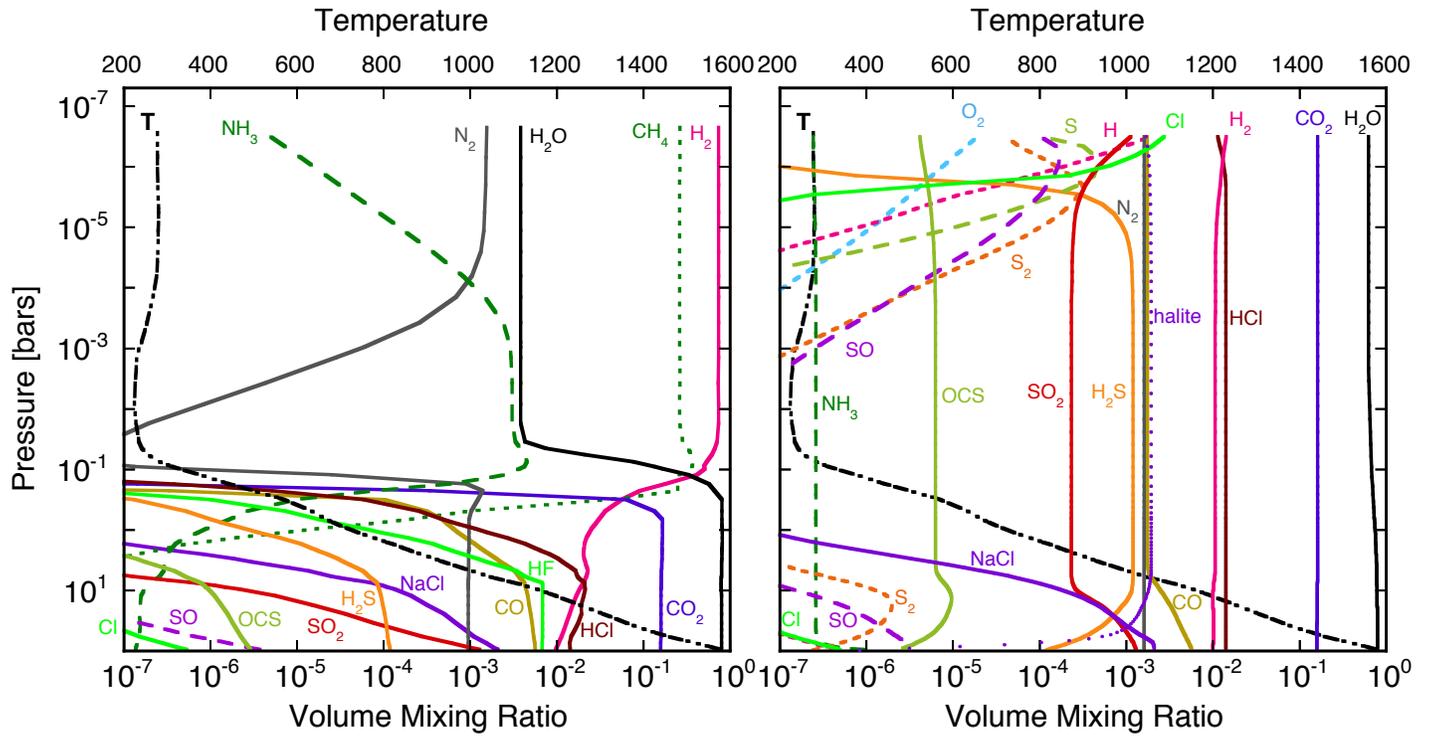}
\caption{Same as Figure~\ref{abunds_cc} for the Bulk Silicate Earth case. Due to quenching of CH$_4$ and NH$_3$ reactions, these species are no longer important in this scenario (left panel). The atmospheric composition will resemble more closely the one for the continental crust case, dominated by water and CO$_2$. However, more sulfur species are produced by equilibrium chemistry in the lower atmosphere than in the continental crust case, and are subsequently mixed in the higher layers. \label{abunds_bse}}
\end{figure}

 The righthand panels in Figures \ref{abunds_cc} and \ref{abunds_bse} show the results of the photochemistry calculation for a 1600~K surface and a 100 bar atmosphere, for atmospheres that at the surface are in equilibrium with CC or BSE, 
respectively. Two things stand out: (i) vertical mixing is very important in these 
atmospheres and (ii) photochemistry is important at the top.
The kinetically computed atmospheres in the infrared will broadly resemble the emission from the atmospheres in thermodynamic equilibrium (TE) at $\sim$1000~K (5 bars). Several species are seen to mix to high altitude, in abundances far in excess of the predictions of TE. Interesting examples include H$_2$S and other sulfur species in the BSE atmosphere, O$_2$ and more SO$_2$ in the CC atmosphere. Other molecules that TE predicts might be abundant are barely seen (CH$_4$ and NH$_3$ are the most important); their absence was expected from quenching as discussed above.

A notable difference between the equilibrium chemistry code and the kinetics code is that the kinetics code predicts that sulfur species should extend throughout the atmosphere. The high abundance of SO$_2$ in the CC case is especially noteworthy. The different sulfur species conform roughly to their relative TE 
abundances at 1000~K, although the absolute abundances are much greater. This is likely to be correct in a general sense; the TE assumption that sulfur species from cool parts of the atmosphere will condense as CaSO$_4$ (say) at the surface is not very realistic.
The kinetics code also predicts that HCl vapor and NaCl crystals (halite) would be mixed to high altitudes. These predictions are more likely to be artifacts,
because the model does not yet account for HCl and NaCl dissolving in 
water clouds. On the other hand the HCl abundance is so high that it may exceed what 
can be accommodated by clouds. HF, which we have not computed, is also likely to stay in the gas phase to high altitudes given the high strength of the HF bond.

The detailed kinetics predictions in Figures \ref{abunds_cc} and \ref{abunds_bse} for CH$_{4}$ and NH$_{3}$ agree well with predictions made with the quenching approximation, described in Section~\ref{sec:quench}. Since most other relevant species are sulfur compounds, identifying the relevant reactions and the corresponding timescales for these species may make it faster and easier to explore the parameter space of possible atmospheric compositions over magma oceans.

\begin {figure}[h]
\begin{minipage}{\linewidth}
      \centering
\begin{minipage}{0.35\linewidth}
\hspace{-0.5\linewidth}
   \includegraphics*[width=1.5\linewidth,angle=0]{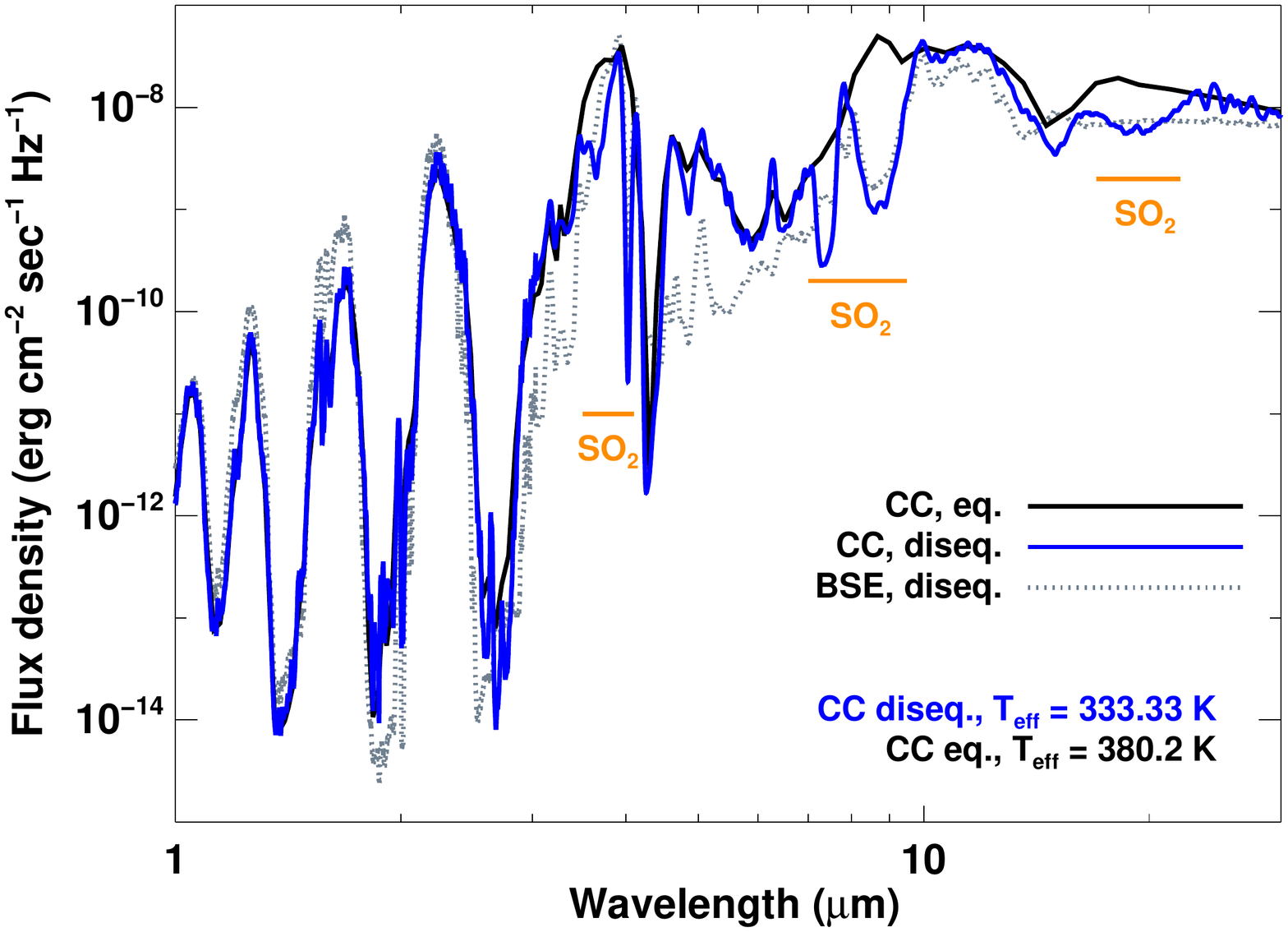} 
 \end{minipage}
      \begin{minipage}{0.35\linewidth}
\includegraphics*[width=1.5\linewidth,angle=0]{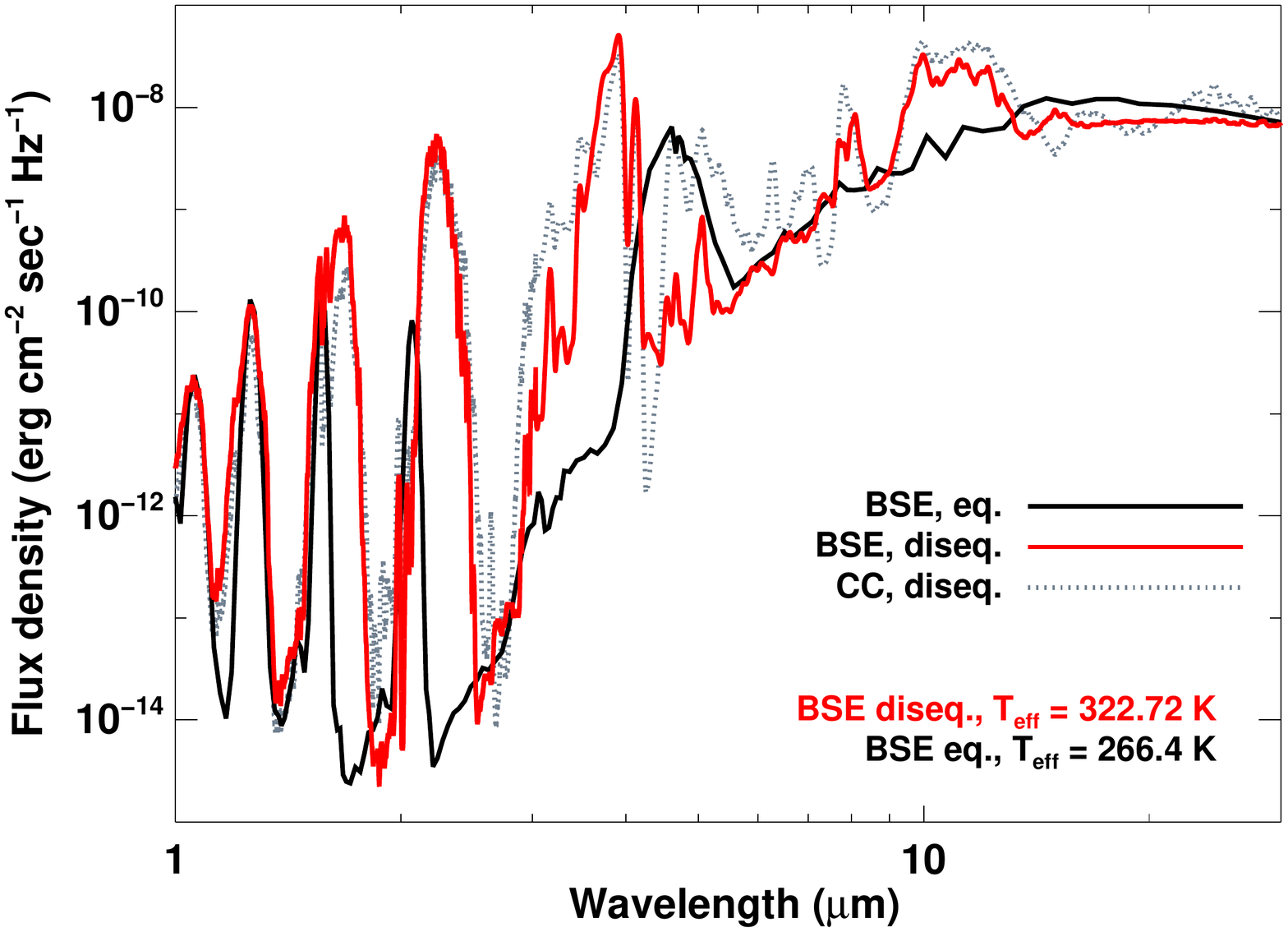} 
\end{minipage}  
   \end{minipage}
\caption{Estimated effects of disequilibrium chemistry on the output spectrum and effective temperature for the CC (left) and BSE (right) cases with $p_{\rm surf}=100$~bar, $T_{\rm surf}$=1600~K, and $g=10$~m~s$^{-2}$. The original models are shown in black, and the models with the disequilibrium abundances are shown in blue (CC) and red (BSE). These correspond to the abundance profiles shown in the right panels of Figures \ref{abunds_cc} and \ref{abunds_bse}, respectively. For a direct comparison, the new BSE and CC models are also reproduced with dotted gray lines on the left and right panels, respectively. The disequilibrium abundances, given the same $T-p$ profile, lead to changes of about 50~K in the effective temperatures. \label{noneq}}
\end{figure}

To explore the impact disequilibrium chemistry might have on the emergent spectrum, we computed the high-resolution spectra (Figure~\ref{noneq}) for the BSE and CC  $p_{\rm surf}=100$~bar and $g=10\,\rm m\,s^{-2}$ cases, using the disequilibrium chemical abundances to replace the equilibrium values. The original models are shown in black, while the alternate composition is shown dotted on both plots, for comparison (CC vs. BSE). Since we do not converge a new thermal profile, the models now have a new emergent flux and $T_{\rm eff}$, which are also indicated on the plots. Especially for the BSE atmosphere (right panel), the spectra show significant differences in spectral morphology, thus the effect of disequilibrium chemistry in these atmospheres would be discernible in even low resolution spectra. For the CC case, the SO$_2$ spectral features become even more prominent. As apparent from the right panel, after taking into account the chemical disequilibrium in the atmosphere, the spectra for the BSE and CC compositions become similar, due to the predominance of water and CO$_2$ in both cases. The BSE model is distinguishable in this case by additional CO absorption.

We stress that the photochemical models are not yet fully self-consistent radiative-convective atmosphere models since they are computed for a fixed atmospheric structure. The computed effective temperature of the altered models is larger in the BSE case without the methane and ammonia opacity, and lower for the CC atmosphere due to extra SO$_2$ opacity. A fully self-consistent model would have to properly account for the effect of mixing as the model converges which is beyond the scope of this preliminary investigation. The removal of CH$_4$ and NH$_3$ from the CC models does not yield significant spectral changes, since the abundances of these compounds were still low in the original models. It is interesting to note, however, that the spectral signatures of HF, HCl, and SO$_2$ should remain present regardless of assumptions about mixing as the spectral features for these molecules are formed in the hotter parts of the atmosphere which are expected to be in chemical equilibrium. In fact, for the CC composition the SO$_2$ features become some of the most prominent, even for 100~bar atmospheres (Figure~\ref{noneq}, left), due to the increase in total column for this important absorber. 

\section{Discussion}

\subsection{Evolution timescales\label{sec:evol}}

The opacity of the atmosphere will be one of the determining factors for the time evolution of the planet's temperature after the giant impact. Understanding this cooling process is also important for characterizing the formation of the Earth-Moon system, which is still subject to many uncertainties in the current models. As shown in Figure~\ref{teff}, even when the surface is at 2200~K, due to the presence of greenhouse gasses in the atmosphere, the planet only radiates at an effective temperature of 500-750~K for the CC, or 450-800~K for the BSE, depending on surface pressure. This limits the amount of heat that can be lost by the planet per unit time. 

We consider the generic case of an Earth-like planet after a giant impact that does not form a Moon. We follow the evolution in time of this planet, starting with a hot molten magma planetary surface at a temperature of 2400~K, which gradually cools and solidifies. The post-impact atmosphere is assumed to have a surface pressure of 100~bar, which remains constant until the planetary mantle completely solidifies. The liquidus is set at 1800~K (above which the mantle is purely liquid), while the solidus is set at 1400~K (below which the mantle is completely solid). In between these two temperatures, the mantle is only partially solidified. The heat stored in the mantle at any given temperature is a function of the heat capacity of the mantle and the fraction of material in liquid state. Assuming a convective interior, the energy flow from the mantle to the surface is governed by the Rayleigh number of the fluid, a function of temperature, viscosity, thermal diffusivity and expansion coefficient. The total energy radiated by the planet is equivalent to that of a blackbody with a temperature equal to $T_{\rm eff}$ (which also includes insolation). The full derivation of the cooling process, as well as the relevant quantities are given in the Appendix. The particular case where the giant impact results in the formation of a satellite such as the Earth's Moon, considering the tidal interaction and orbit evolution of the two bodies, will be adressed in a future paper.

Figure \ref{evolution} shows the results of our calculations for an Earth-like planet. The interior, surface and radiating (effective) temperatures are shown in red, blue, and black, respectively. The BSE model with $p_{\rm surf}=100$~bar (left) has a stronger greenhouse and spends more time above the liquidus. Even though in such a scenario the surface of the planet stays hot for a longer time, the low effective temperature will make it harder to detect than a planet with a more transparent atmosphere (such as the CC model). The sharp transition in the surface temperature marks the point where the mantle begins to solidify, and its viscosity becomes large (1560~K, see Appendix). As the interior temperature drops below this value, the heat transfer from the interior to the surface becomes small, and the surface temperature is solely determined by the energy balance with the solar radiation, therefore does not change, giving rise to the sharp transition in Figure \ref{evolution}, the breadth of which is determined by the relatively tiny heat capacity of the atmosphere. This analysis shows that the models discussed in this work are applicable to timeframe spanning $\sim10^{7}$~years after the planet has undergone a giant impact. The planet will be in a hot state and therefore more easily detectable for a time period of $\sim10^{5}$~years after the impact. Finally, we note that in view of the discussion regarding disequilibrium chemistry, the BSE atmosphere will be more transparent, while the CC atmosphere will be more opaque than used in this calculation. We expect that this change will result in a shorter (longer) cooling timescale for the BSE (CC) cases, in which case the evolution for the two atmospheric compositions will resemble each other more closely. Nevertheless, the BSE equilibrium composition can be considered as a relevant case for a planet undergoing a runaway greenhouse stage during its cooling process.

\subsection{Observing post-giant-impact Earths\label{sec:detect}}

\subsubsection{Detection}

As discussed in Section \ref{sec:mags}, despite their high surface temperatures, post-impact Earths will have relatively low effective temperatures; this will make them extremely challenging to observe. The observational limit is primarily due to the ability to suppress the host star's light with adaptive optics and a coronagraph. The first ``extreme" adaptive optics systems on 8-m class telescopes (GPI, SPHERE) will achieve $\sim10^{-7}$ contrast at angular separations greater than 100 mas; for a Sun-like star observed in $H$-band, this corresponds to an absolute $H$ magnitude of $\sim$ 21. This means that the hottest models in this study ($p_{\rm surf}=10\, \rm bar$ and $T_{\rm surf}=2200\, \rm K$) will be at the edge of the reachable contrast limit, in contrast to the models of \citep{Miller-Ricci:2009}, which estimate a contrast between $10^{-6}$ and 10$^{-7}$ for a 1$M_{\oplus}$ planet with similar surface pressure and temperature, and place Earth-mass planets with $p_{\rm surf}=100\,\rm bar$ and $T_{\rm surf}=1500\,\rm K$ at the limit of detectability.

For comparison, in Figure~\ref{colormag} we show the observed brown dwarfs (gray circles) and directly imaged exoplanets (purple circles). The brightest of our models lie at about 6 magnitudes fainter in $H$-band than the currently imaged extrasolar planets, namely 2M1207b \citep{Patience:2010} and the planets around HR~8799 \citep{Skemer:2012, Marois:2010}. However, following the trend of $H$-band magnitude with surface pressure (lighter vs. darker symbols), post-impact Earths with tenuous atmospheres ($\sim$1~bar) would fall several magnitudes closer to the HR~8799 planets, due to the hot surface being less obscured. If at an appropriate angular separation such planets would be in the detectability range of the 8-m class telescopes, but over a shorter timescale, since they would cool much faster than a young giant planet. Such observations would test the theories of volatile-poor accretion during terrestrial planet formation, as these nearly-volatile-free hot Earths would be the brightest among similar size planets for a brief period of time.

A thirty-meter class telescope with extreme adaptive optics will be capable of an order of magnitude higher contrast; \citet{Macintosh:2006} estimates that this class of telescope will be capable of 10$^{-8}$ contrast at 1.65 \micron\ at angular separations of tens of mas. For a Sun-like G2 dwarf, this corresponds to a $H$-band absolute magnitude of 23.3; for a fainter M1 dwarf, the limiting magnitude is 25.3. Figure~\ref{colormag} shows these magnitude limits; many of the hot post-impact Earths fall above these lines, suggesting that they will be observable with a thirty-meter class telescope, especially if cooler stars are targeted.

Following \citet{Stern:1994}, the average number of stars we would need to search to find one post-giant-impact Earth-like planet is $N_{*}=1/N_t/(t_{acc}/t_{hot})$, where $N_t$ is the number of terrestrial planets per star, $t_{acc}$ is the timespan of the giant impact era during terrestrial planet formation, and $t_{hot}$ is the length of time the planet stays hot during the entire period of giant impacts. The latest estimates \citep{Bonfils:2013,Kopparapu:2013} give an occurrence rate of about 0.4-0.6 terrestrial planets in the habitable zone, per star.  According to terrestrial planet formation models \citep{Wetherill:1992,Agnor:1999}, $t_{acc}$ is in the range 40-100~Myr. The time the planet is hot after each giant impact is about ~$3\times10^5$~yrs (see Figure~\ref{evolution}). \citet{Stewart:2012} find that it takes on average between 10 and 16 giant impact to make an Earth-like planet. Therefore, the total $t_{hot}$ in the best case scenario will be $16\times3\times10^5$~yrs. Therefore, the number of young stars to be observed in order to find a post-giant impact Earth-like planet is about 14 in the best-case scenario. This number becomes $\sim40$ in a not-so-favorable region of the parameter space. 

There are a number of nearby ($< 50\,\rm pc$) young stellar clusters in the age range where giant impacts would be expected among the young terrestrial planets. Examples include the Tucana-Horologium association (age 10$-$30 Myr, with about 50 stars at 50~pc \citep{Zuckerman:2001}) and the closer AB Dor moving group (age $\sim 50\,\rm Myr$, about 30 stars at 20 pc \citep{Zuckerman:2004}). The $\beta$ Pic moving group, the TW Hydrae association, and the $\eta$ Cha cluster are all younger and include more stars. There are clearly a sufficient number of nearby young stars to motivate a survey for  post-giant impact planets. This estimate is in agreement with previous calculations by \citet{Miller-Ricci:2009,Jackson:2012}, who find that about 10\% of the stars could host such a post-giant impact planet. One of the goals of future surveys will be precisely to constrain the number of stars hosting such post-giant impact planets (i.e. $N_*$ as defined above), and narrow the parameter space currently in use by the terrestrial planet formation models. We emphasize that our estimate takes into account only terrestrial planets the habitable zone. The number of possible detections could increase by factors of a few if we consider planets outside this zone, as well as non-Earth-like planets.

\subsubsection{Moons and Disks}

Anther direct consequence of a giant impact is the scattering of debris, with the formation of a circumstellar debris disk, and the possible formation of a moon. Obseving a newly-formed moon is unlikely. The probability of forming a moon is $\sim8$\% \citep{Elser:2011}. Although the moon is formed hot and potentially much brighter due to the lack of atmosphere, its surface cooling time is comparatively short relative to the planet, about 400~yrs \citep[][e.g.]{Pritchard:2000}. By the same argument as above, the number of stars one would need to observe in order to find such a moon would be on the order of $10^4$. It is clearly more likely that future surveys will detect the post-giant impact planet, even with a cooler $T_{eff}$, rather than the moon, which only forms in 8\% of the cases.

However a giant impact will not only melt the target but will likely create a disk of impact-generated debris surrounding the star in the orbit of the target planet. Such a disk might both be a signature of the impact and a hinderance for efforts at direct imaging in the near-infrared. The degree to which this is true depends on the evolution of the disk, including effects of Poynting-Robertson drag and gap clearing by the planet itself. Indeed the observational signature of an impact-generated debris disk is an area of active research, as part of the extrasolar zodiacal light (exozodi) issue, which affects planet detectability in general \citep[e.g.][]{Roberge:2012, Turnbull:2012}. A detailed study of the signature of such a disk is well beyond the scope of this paper but we present a few notional calculation to estimate its effect to provide a comparison to the estimated flux of the planet. 

\begin {figure}[h]
\hspace{-0.1\linewidth}
\includegraphics*[scale=0.7,angle=0]{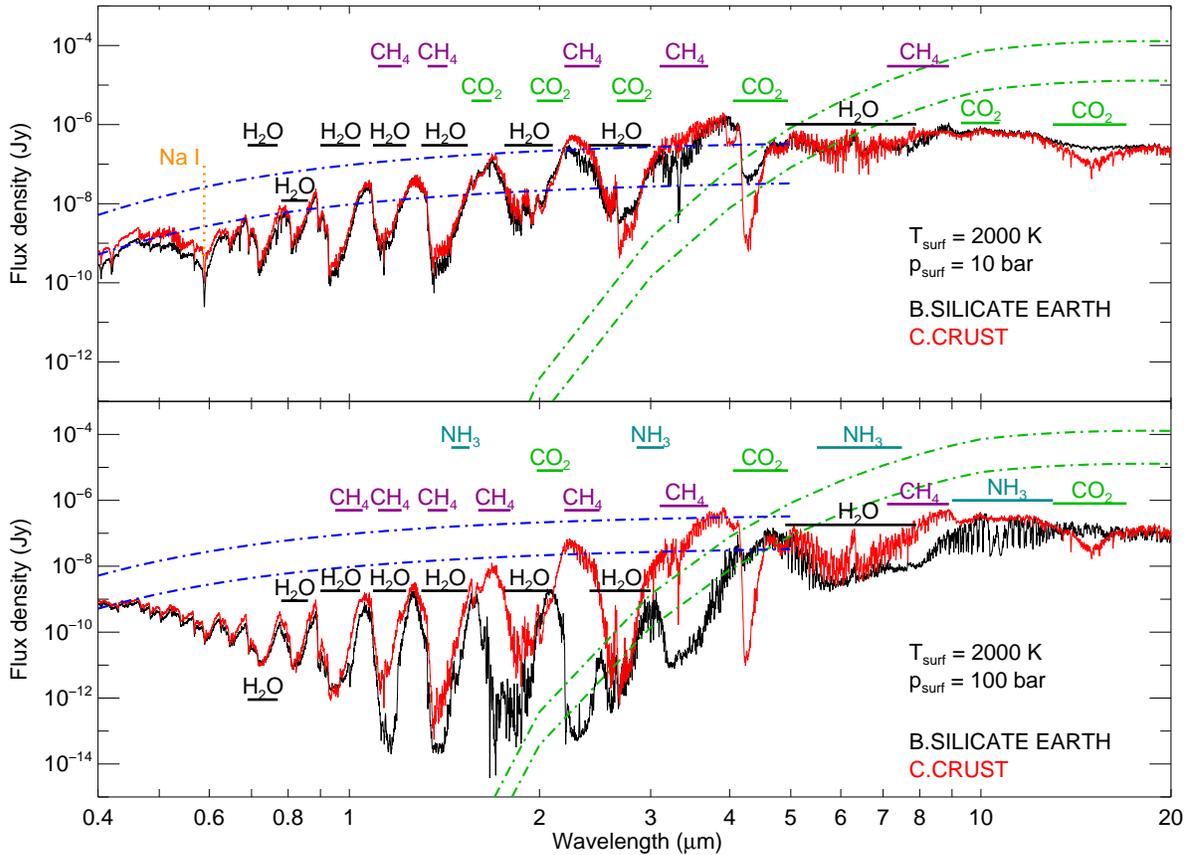}
\caption{Comparison between the high-resolution spectra for the Continental Crust (red) and the Bulk Silicate Earth (black) models, with $T_{\rm surf}=2000$~K, $g=10$~m~s$^{-2}$.  For the models with $p_{\rm surf}=10$~bar (top panel), we note the stronger CO$_{2}$ features (green) for the Continental Crust case, and the contribution of CH$_{4}$ to the Bulk Silicate Earth model. The \ion{Na}{1} D lines (orange) are present in both cases. The bottom panel shows the same comparison for $p_{\rm surf}=100$~bar. While the features for the Continental Crust model do not change considerably, the Bulk Silicate Earth model becomes dominated by CH$_{4}$, and NH$_{3}$ replaces CO$_{2}$ as a dominant contributor. The estimated contribution of a post-impact debris disk is shown by the dot-dashed lines. The blue and green curves represent the reflected and thermal dust emission, respectively, per resolution element of a 30-m telescope. The two sets of curves correspond to total luminosities of the disk of 1 and 10 zodis. The flux density is calculated at the observer, assuming a system 100 light years away, and a 0.8 AU wide disk at 1~AU from the star. \label{cc_bse_comp}}

\end{figure}

In general the direct detection of any planet at optical wavelengths will require a space-based mission, so we are primarily concered with scattered light and thermal emission from the disk at infared wavelengths.
The post-giant impact debris will cool to a temperature of $\sim$278~K (at 1~AU) on a $100-1000$~yrs timescale \citep{Jackson:2012, Kains:2011}. Thermal emission from such a disk will peak in the mid-infrared. Depending on the size of the largest fragment, the post-impact debris disk luminosity $10^5-10^6$~years after the impact can be about $10^{-3}-10^{-7}$ that of the star. The lower limit is equal to the unit of 1 zodi. Since  currently planned missions aim for levels of 10 zodi or less for detecting Earth-like planets, we use the 1-10 zodis range as the scaling for the integrated luminosity of the disk. Levels as high as 100 zodis would also be acceptable, coupled with a significant increase in exposure time \citep[e.g.][]{Turnbull:2012}. Figure~\ref{cc_bse_comp} shows that the radiation from the post-giant impact planet will exceed the integrated thermal radiation from the dust (green dot-dashed line) at wavelengths shorter than about $\sim$4~$\mu$m. With a 30-m class telescope the disk will be resolved, and the light from the disk will be spread out a large number of pixels, increasing the contrast between the disk and the planet. The flux density from the debris disk in Figure~\ref{cc_bse_comp} is calculated per resolution element (equal to a diffraction-limited circle at all wavelengths), assuming the dust has a circular distribution 0.8~AU-wide, centered at 1~AU from the star. The post-giant impact system is placed 100 light-years (about 30 pc) away from the observer.

Virtually nothing is known about the exozodi scattering properties of disks around young stars, since most debris disks are discovered in the far-IR. Assuming a constant optical albedo of 0.12 (similar to the Moon's), and using the same luminosity scaling as above (1 and 10 zodis), the scattered light from the disk is shown in blue in Figure~\ref{cc_bse_comp}. In this case, the scattered light from the disk can be larger than the emission from the planet in the visible, but we find that the radiation from the 10-bar atmosphere could be stronger than a 10-zodi background in the 2-4~$\mu$m range. It is apparent from the same figure that the water windows below 4~$\mu$m would offer the best contrast for detecting the post-giant impact planets. The detection of infrared excess at $10-20 \mu$m can signal the presence of a debris disk, and such stars can be then followed up at shorter wavelengths ($1-4 \mu$m) with high resolution adaptive optics in order to detect potential post-giant impact planets. The possible complication from a brighter debris disk can be dealt with observationally. If the planet is imaged directly, some light from the disk will fall into the planet's PSF, but this can be efficiently subtracted as a background, using the SED measured using the extra pixels occupied by the disk. In this case, the contribution from the disk will not affect the atmospheric signatures, but will likely add noise to the spectrum. We refer the reader to \citep{Roberge:2012,Turnbull:2012} for a more rigorous calculation and in-depth description of these issues.

\section{CONCLUSIONS}

We have investigated the atmospheres of Earth-like planets following a giant impact event.
As discussed in more detail by \citet{Schaefer:2012}, the chemical compositions considered, BSE and CC, characterize a class of rocky planets around stars of solar or near-solar metallicity, with a differentiated metal + FeS bearing core and a silicate exterior, with all Fe as Fe$^{2+}$ and Fe$^{3+}$ in minerals. These post-giant impact atmospheres are constituted of vapors in equilibrium with the magma ocean at the bottom, and gases rising through turbulent mixing at altitudes where the temperature drops below 900-1000~K. The main constituents, dominating the energy transport through the atmosphere, are water and CO$_2$. We find that the post-impact chemistry also produces characteristic molecules, namely HCl, HF, alkali halides, SO$_2$, and other sulfur-bearing species that are not expected in cooler atmospheres. By contrast, anhydrous worlds have no volatiles, hence no HF and HCl \citep{Schaefer:2009}. We also note the presence of significant amounts of Na and K in these atmospheres, with potentially detectable \ion[1][Na] lines for thinner (10 bar) atmospheres. The spectral signatures of these species can thus be considered as markers for post-giant impact Earth-like planets. However, the spectral resolution and contrast needed for such observations is still beyond the capabilities of current instrumentation for exoplanet characterization. Nevertheless the near-infrared colors of such worlds are expected to be notably redder than young giant planets and may thus be at least tentatively identified on this basis. 

The blanketing effect of the atmosphere obscures the hot surface of post-impact planets at infrared wavelengths. For surface temperatures between 1000 and 2200~K, the corresponding effective temperatures are found to be only as high as $\sim$700~K for an Earth-like planet with $p_{\rm surf}=100$~bar, and possibly as low as 265~K when the planet is completely in a runaway greenhouse state. These low effective temperatures, as well as the addition of more opacity sources which diminish the escape of radiation through the H$_2$O and CO$_2$ windows, make the post-giant impact planets in these models less bright than previously thought \citep[e.g.,][]{Miller-Ricci:2009}. Due to the added opacity sources, the brightness temperature in the water windows is decreased by factors of 2-3. This translates in flux densities that are far fainter in the spectral windows than predicted by \citet{Miller-Ricci:2009}, depending on wavelength. In the case of thin, tenuous atmospheres ($\sim$1~bar), or larger surface gravities, some of these planets may still be detectable with 8-m class telescopes. However, for the higher surface pressure cases discussed in this paper, even the hottest planets will only be accessible to 30-m class telescopes. 

We estimate, based on the heat transport from the planet's interior and through the atmosphere, that these post-giant impact plantes will stay hot on the order of one million years, and may undergo several such impacts during their formation. As an outcome of a giant impact, the hot planet is definitely more likely to be observed than a potential young moon. In the 1-4 $\mu$m wavelength region sufficient contrast may also be achieved to distinguish such a planet from a surrounding debris disk. Given the observational capabilities, it is thus plausible that some planets will be found in this state by direct imaging with coronagraphs on large telescopes. Based on the expected frequency of terrestrial planets, we estimate that one such post-giant impact planet may be discovered for every 10-50 stars surveyed in young stellar clusters.

Our models do not yet include opacities from clouds and aerosols, which may further diminish the predicted brightness temperatures in the infrared. Disequilibrium chemistry also proves to be important for these atmospheres, and a self-consistent approach, including vertical mixing, photochemistry, and clouds into the calculation of the radiative-convective atmosphere are subjects for future work.

\acknowledgments
This work was supported by the NASA Origins program. B.F. was supported by the NSF Astronomy Program, the NASA EPSCOR Program, and by NASA Cooperative Agreement NNX09AG69A with the NASA Ames Research Center.

\appendix
\section{Appendix}

We describe magma ocean cooling by taking into account convective heat
flow through the mantle, the heat capacities of the mantle and core, the heat of fusion when the mantle congeals, and the thermal blanketing effect imposed by the atmosphere.
Plausibly important factors that we do not include here are high heat flow from a superheated core, tidal heating by a nearby moon or moons, tidal heating by the sun and, if very early, heating by decay of $^{26}$Al. We will address tidal heating by a moon elsewhere.  
  
For simplicity we treat the interior as characterized by a single potential temperature $T_{\rm i}$, the temperature that any parcel in an adiabatic mantle would have were it brought to the surface, while the surface temperature itself is $T_{\rm surf}$. 

We describe silicate freezing by three temperatures: a liquidus $T_{\rm liq}$ hotter than which the mantle is fully molten; a solidus $T_{\rm sol}$, colder than which the mantle is fully solid; and a critical temperature $T_{\rm crit}$ in between where the rheology of the material changes from that of a solid with melt percolating through it to a liquid in which solids are suspended \citep{Abe:1997,Solomatov:2007}. We take $T_{\rm liq}=1800\,\rm K$ and $T_{\rm sol}=1400$K, and assume that the melt fraction $\phi$ is linear in $T_{\rm i}$,
\begin{equation}
\phi = \begin{cases} 1, &\quad T_{\rm i}>T_{\rm liq} \\
\left(T_{\rm i}-T_{\rm sol} \right) /\left( T_{\rm liq}-T_{\rm sol} \right), &\quad T_{\rm sol}<T_{\rm i}<T_{\rm liq} \\
0 , &\quad T_{\rm i}<T_{\rm sol} 
\end{cases}
\end{equation}

For $T_{\rm liq}<T_{\rm i}<T_{\rm sol}$, we approximate the heat capacity of our post-giant-impact planet with that of the Earth,
$CM_{\oplus} = C_{\rm v} M_{\rm man} + C_{\rm c} M_{\rm core} = 6.2\times 10^{34}$ ergs~g$^{-1}$~$K^{-1}$, in which
the heat capacity of silicate is $C_{\rm v}=1.2\times 10^7$ ergs~g$^{-1}$~$K^{-1}$ and for the iron core is $C_{\rm c}=6.5\times 10^6$ ergs~g$^{-1}$~$K^{-1}$. 
The iron core does not freeze on the time scale of interest.
While silicates freeze, the effective heat capacity of the Earth is 
 \begin{equation}
 C^{\prime}M_{\oplus} = CM_{\oplus} + {Q_{\rm m}M_{\rm man}\over T_{\rm liq}-T_{\rm sol}} = 1.0 \times 10^{34} {\rm~ergs/K}
 \end{equation}
where the heat of fusion is $Q_{\rm m}=4\times 10^9$ ergs~g$^{-1}$, and $M_{\rm man}$ is the mass of the mantle, $M_{\rm man}=4.1\times 10^{27}$~g. 

Following \citet{Solomatov:2007}, we treat mantle cooling by parameterized convection.
In this framework, heat flow is related to the size and viscosity of the interior and the temperature gradient across the surface boundary layer through the Rayleigh number $Ra$. 
\begin{equation}
\label{Heat_flow}
F = C_0 {k_{\rm c} \left(T_{\rm i} -T_{\rm surf} \right) \over L}  Ra^n
\end{equation}
where $C_0=0.089$ is a constant \citep{Solomatov:2007}, 
$k_{\rm c}=3\times 10^5$~ergs~cm$^{-2}$s$^{-1}$K$^{-1}$ is the thermal conductivity, and $L$ a characteristic distance usually identified with the thickness of the mantle ($L=2.8\times 10^8$ cm). The power $n=1/3$ in soft turbulence \citep{Solomatov:2007}.
The Rayleigh number describes the ratio of bouyancy to viscosity,
\begin{equation}
\label{Rayleigh}
Ra = {\alpha_{\rm v} g \left(T_{\rm i} -T_{\rm surf} \right) L^3 \over \kappa \nu}
\end{equation}
In Eq \ref{Rayleigh}, $\alpha_{\rm v} = 2.4\times 10^{-5}$ K$^{-1}$ is the volume thermal expansivity, $g$ the gravity, $\kappa = 0.01$ cm$^2$s$^{-1}$ is the thermal diffusivity ($=k_{\rm c}/C_{\rm v}/\rho$), and $\nu$ the kinematic viscosity (cm$^2$s$^{-1}$).

Viscosity is a strong function of temperature. Following \citet{Abe:1997,Solomatov:2007,Lebrun:2013}, we divide viscosity regimes into a rheological liquid and a rheological solid. In the rheological liquid, Solomatov parameterizes viscosity by 
\begin{equation}
\label{liquid_viscosity}
\nu = 6\times 10^{-4} \exp{\left({4600\over T_{\rm i}-1000}\right)} \left({1-\phi_{\rm crit}\over \phi -\phi_{\rm crit}}\right)^{2.5} 
\end{equation}  
This expression for $\nu\rightarrow\infty$ as $\phi\rightarrow\phi_{\rm crit}$; i.e., it is singular at the critical value $\phi=\phi_{\rm crit}$.
In the rheological solid, 
\begin{equation}
\label{solid_viscosity}
\nu = 1\times 10^{19} \exp{ \left( {T_{\rm sol}-T_{\rm i}\over 58} \right) } \exp{\left(-\alpha_{\phi} \phi \right)} 
\end{equation}
where $26<\alpha_{\phi}<32$ \citep{Solomatov:2007}.
\citet{Abe:1997} places the boundary at $\phi = \phi_{\rm crit} = 0.4$; the precise value of $\phi_{\rm crit}$ is not very important. For $\phi_{\rm crit} = 0.4$, the critical temperature for the rheological transition is $T_{\rm crit}= 1560 {\rm ~K}$.

Equations \ref{Heat_flow} and \ref{Rayleigh} can be combined to give
\begin{equation}
\label{combined}
F = C_0 k_{\rm c} \left(T_{\rm i} -T_{\rm surf} \right)^{(1+n)} L^{(3n-1)} \left( {\alpha_{\rm v} g \over \kappa \nu}   \right)^{n}
\end{equation}
It is usual to take $n=1/3$, in which case $L$ cancels out. This gives a system in which the heat flow is determined only by the properties of the boundary layer, which is often regarded as the physically meaningful case.

\begin {figure}[h]
\centering
\includegraphics*[scale=0.6,angle=0]{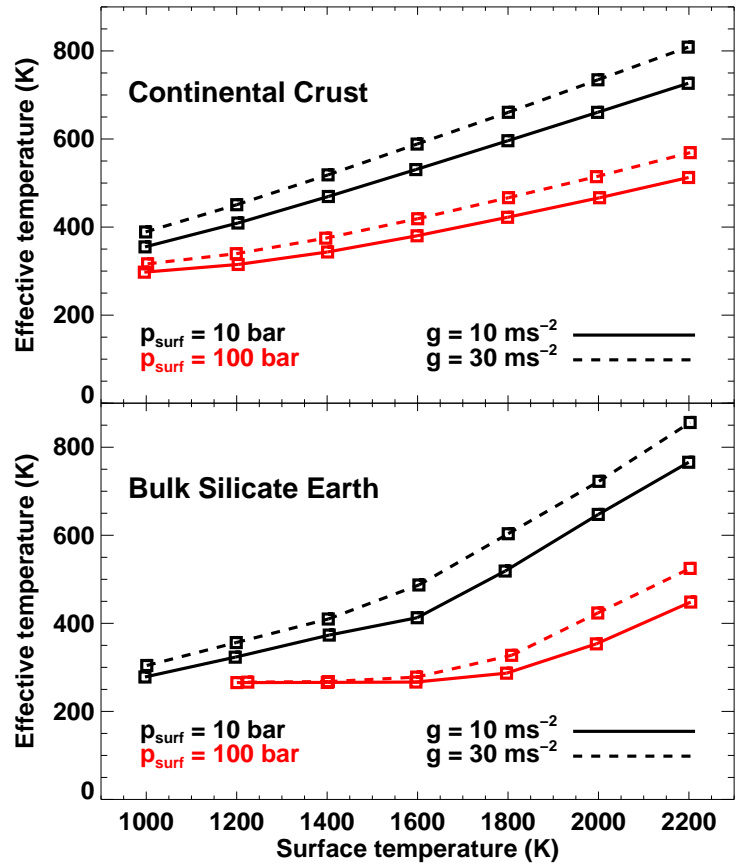}
\caption{Effective temperature variation as a function of surface temperature for the Continental Crust (top panel) and Bulk Silicate Earth (bottom panel) models. At a give surface temperature, the effective temperature is governed by both $p_{\rm surf}$ and $g$, with larger variations for the Bulk Silicate Earth models. The flattening of the $T_{\rm eff}$ curves for the Bulk Silicate Earth models with $p_{\rm surf}=100$~bar reflects the optically thick regime where $T_{\rm eff}$ is determined by the amount of insolation, and not by the surface temperature. Effective temperature values for the models are tabulated in Table~\ref{tab:teff}. \label{teff}}
\end{figure}

The cooling of the planet interior is described by
\begin{eqnarray}
\label{interior cooling}
CM_{\oplus} {dT_{\rm i}\over dt} = -4\pi R_{\oplus}^2 F & \qquad T_{\rm i}>T_{\rm liq}, T_{\rm i}<T_{\rm sol};  \nonumber \\
C^{\prime}M_{\oplus} {dT_{\rm i}\over dt} = -4\pi R_{\oplus}^2 F  & \qquad T_{\rm sol}< T_{\rm i} < T_{\rm liq}. 
\end{eqnarray}
Taking into account the incoming solar flux, $F_{\rm sun}$, and the total heat capacity of the atmosphere, $C_{\rm A}$ (ergs~g$^{-1}$~K$^{-1}$), then the global planetary cooling rate is described by
\begin{eqnarray}
\label{global cooling}
CM_{\oplus} {dT_{\rm i}\over dt} + C_{\rm A} {dT_{\rm surf}\over dt} = -4\pi R_{\oplus}^2 \sigma T_{\rm eff}^4 + \pi R_{\oplus}^2F_{\rm sun}\left(1-A\right)  & \qquad T_{\rm i}>T_{\rm liq}, T_{\rm i}<T_{\rm sol}  \nonumber \\
C^{\prime}M_{\oplus} {dT_{\rm i}\over dt} + C_{\rm A} {dT_{\rm surf}\over dt} = -4\pi R_{\oplus}^2 \sigma T_{\rm eff}^4 + \pi R_{\oplus}^2F_{\rm sun}\left(1-A\right)  & \qquad T_{\rm sol}< T_{\rm i} < T_{\rm liq}
\end{eqnarray}
The net radiative cooling is determined by the effective radiating temperature of the atmosphere, taking into account the sunlight absorbed, where $F_{\rm sun}\approx 1\times 10^6\,\rm ergs\, cm^{-2}\, s^{-1}$ was the solar constant ca.\ 4.5 Ga,
and $A$ is the planetary albedo ($A=0.3$ today). The effective radiating temperature $T_{\rm eff}$ is determined by the thermal blanketing of the atmosphere and the surface temperature $T_{\rm surf}$. The relation between $T_{\rm eff}$ and $T_{\rm surf}$ is derived from fits to the curves in Figure~\ref{teff}, for $1000 < T_{\rm surf} < 2200$~K. These curves are derived for equilibrium chemistry. After taking into account vertical mixing (Section \ref{diseq}), we expect that the curves for the BSE and CC cases will resemble each other more closely.

\begin{figure}[h]
\begin{minipage}{\linewidth}
      \centering
\begin{minipage}{0.4\linewidth}
\hspace{-0.4\linewidth}
   \includegraphics*[width=1.2\linewidth,angle=0]{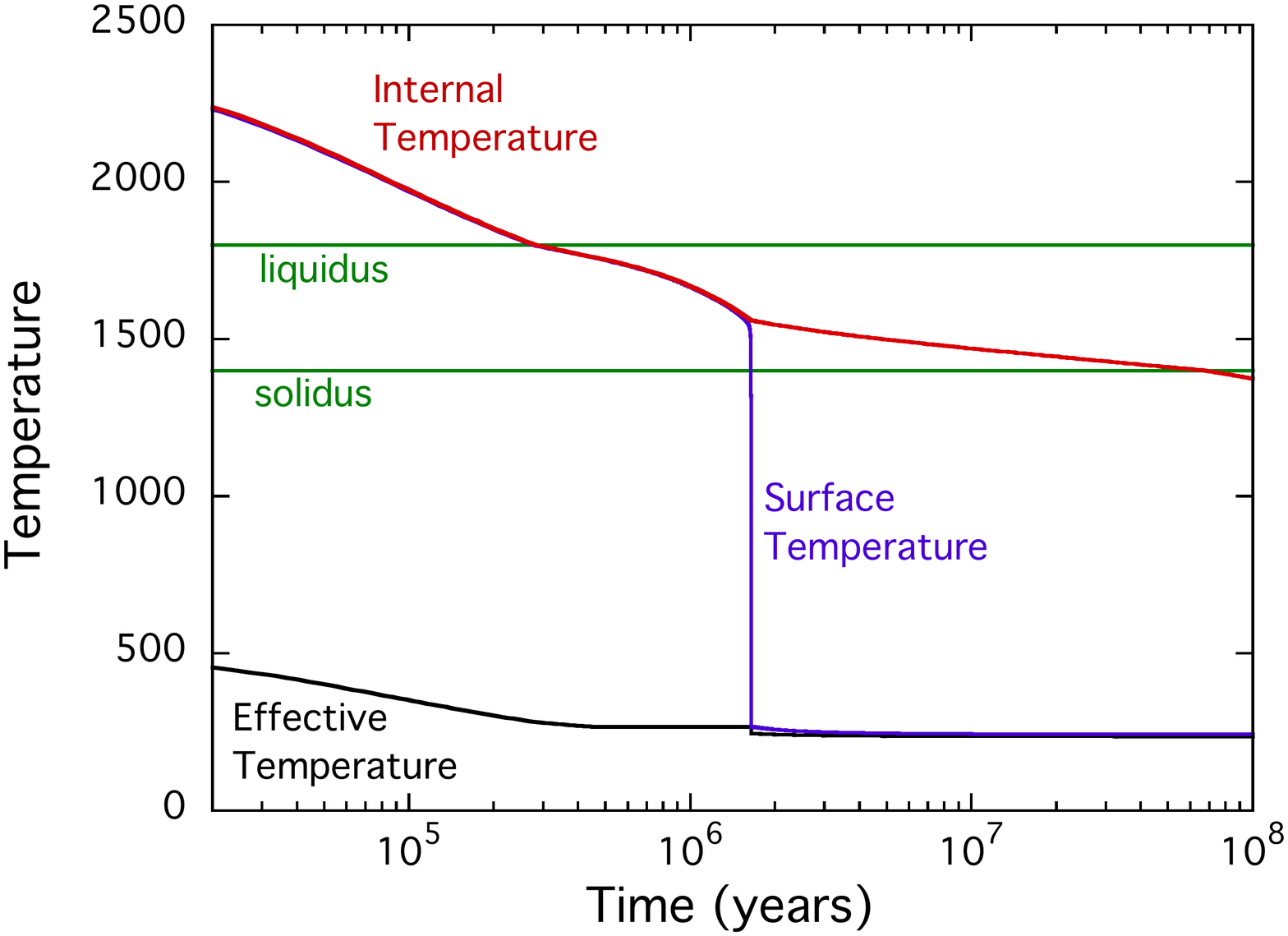} 
 \end{minipage}
 \hspace{-0.07\linewidth}
      \begin{minipage}{0.4\linewidth}
\includegraphics*[width=1.2\linewidth,angle=0]{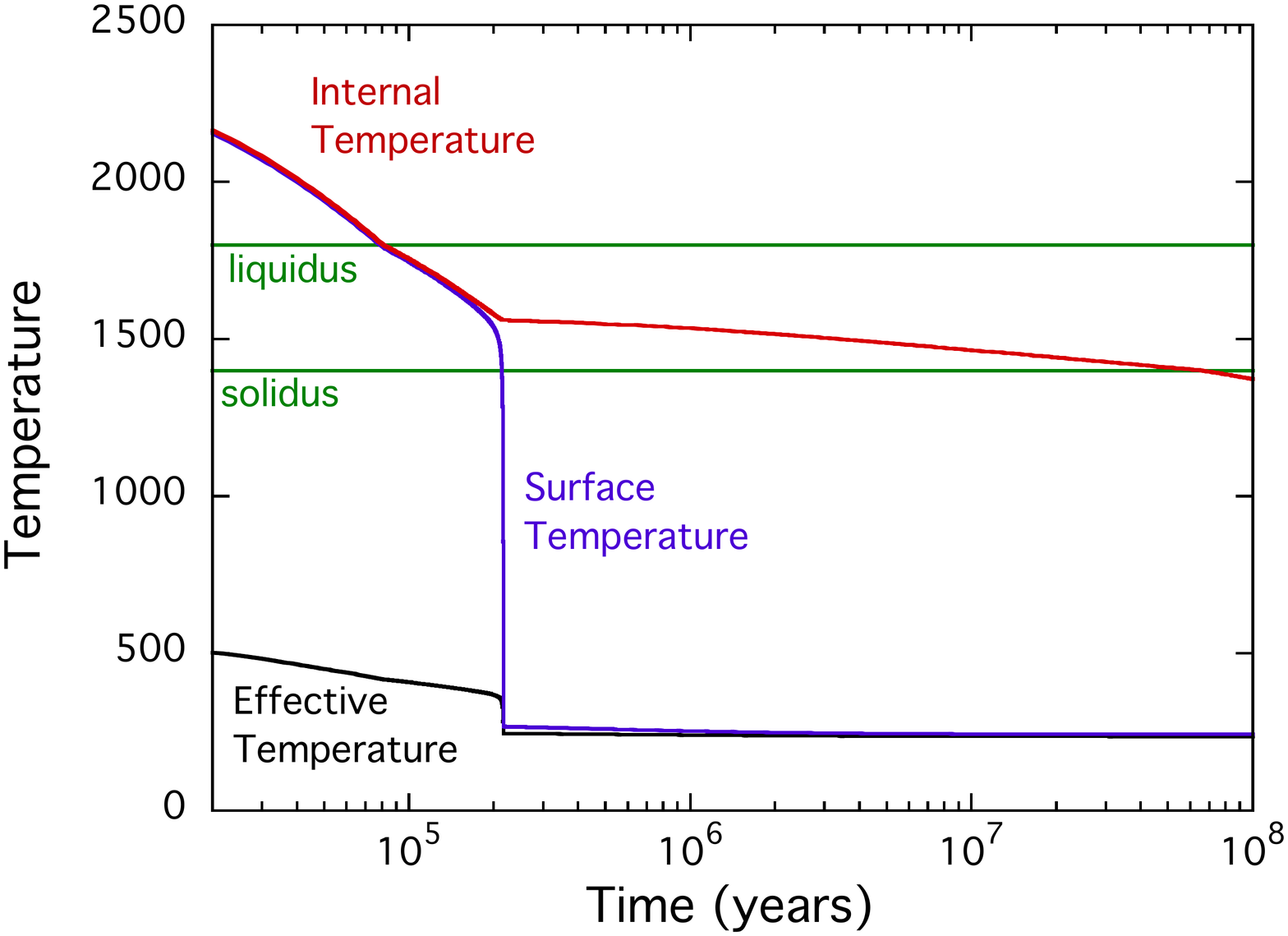} 
\end{minipage}  
   \end{minipage}
   \caption{Thermal evolution after a moonless giant impact in the case of the BSE (left) and Continental Crust (right) 100 bar atmosphere. The BSE experiences a stronger greenhouse effect than the CC. The weaker greenhouse effect gives faster cooling for the Continental Crust case. Internal temperature $T_{\rm i}$, surface temperature $T_{\rm surf}$, and effective radiating temperature $T_{\rm eff}$ are shown. The liquidus and solidus temperature points are indicated in green. The transition from the high heat flow regime to low heat flow (constant $T_{\rm surf}$) takes place when the viscosity of the mantle becomes large, at $T_{\rm crit}=1560$~K (see Appendix). The overall picture presented here closely follows that first discussed by \citet{Abe:1993}. }
   \label{evolution}
\end{figure}

Equations \ref{interior cooling} and \ref{global cooling} are solved for the evolution of $T_{\rm i}$, $T_{\rm surf}$, and $T_{\rm eff}$. Figure \ref{evolution} shows possible thermal evolutions of an Earth-like planet after a giant impact that does not produce a Moon.

\clearpage
\bibliography{references}

\begin{deluxetable}{ccc}
\tablecolumns{3}
\tabletypesize{\footnotesize}
\tablewidth{0pt} 
\tablecaption{\label{tab:comp}Composition of vaporized material.}
\tablehead{Compound & Continental Crust\tablenotemark{a,c} &  Bulk Silicate Earth\tablenotemark{b,c} \\
 & wt. \% &  wt. \%}
\startdata
SiO$_2$ & 64.0 & 45.9 \\
MgO & 2.4 & 37.1 \\
Al$_2$O$_3$ & 14.7 & 4.6 \\
TiO$_2$ & 0.59 & 0.22 \\
FeO & - & 8.2 \\ 
Fe$_2$O$_3$ & 4.9 & - \\
CaO & 4.1 & 3.7 \\
Na$_2$O & 2.9 & 0.35 \\
K$_2$O & 3.1 & 0.03 \\
P$_2$O$_5$ & 0.17 & - \\
MnO & 0.08 & - \\
\hline
\sidehead{Elemental breakdown}
H & 0.045 & 0.006 \\
C & 0.199 & 0.006 \\
N & 0.006 & 0.88e-4 \\
O & 47.20 & 44.42 \\
S & 0.070 & 0.027 \\
F & 0.053 & 0.002 \\
Cl & 0.047 & 0.004 \\
Si & 28.80 & 21.61 \\
Al & 7.96 & 2.12 \\
Fe & 4.32 & 6.27 \\
Ca & 3.85 & 2.46 \\
Na & 2.36 & 0.29 \\
Mg & 2.20 & 22.01 \\
K & 2.14 & 0.02 \\
Ti & 0.401 & 0.12 \\
P & 0.076 & 0.008 \\
Cr & 0.013 & 0.29 \\
Mn & 0.072 & 0.11 
\enddata
\tablenotetext{(a)} {\citet{Wedepohl:1995} }
\tablenotetext{(b)} {\citet{Fegley:2012} }
\tablenotetext{(c)} {\citet{Schaefer:2012} }
\end{deluxetable}

\begin{deluxetable}{ll}
\tabletypesize{\footnotesize}
\tablewidth{0pt} 
\tablecaption{\label{tab:opac}Molecules used for opacity calculations.}
\tablehead{\colhead{Molecule Name} & \colhead{Opacity Source(s)}}
\startdata
 C$_2$H$_2$ & HITRAN'08\tablenotemark{a} with 2011 update \\
 C$_2$H$_4$ & HITRAN'08\tablenotemark{a} \\
 C$_2$H$_6$ & HITRAN'08\tablenotemark{a} with 2010 update; \citet{Lattanzi:2011} \\
 CaH & \citet{Weck:2003a}\tablenotemark{b} \\
 CH$_4$ &  \citet{Brown:2005};\citet{Strong:1993}; \citet{Wenger:1998}\tablenotemark{c}; HITRAN'08\tablenotemark{a} isotopes \\
 ClO & HITRAN'08 \tablenotemark{a,d} \\
 CO & HITEMP'10\tablenotemark{e}; \citet{Tipping:1976} \\
 CO$_2$ &  \citet{Wattson:1986}; \citet{Dana:1992}; HITRAN'08\tablenotemark{a} isotopes\\
 CrH &  \citet{Burrows:2002}  \\
 FeH &  \citet{Dulick:2003}; \citet{Hargreaves:2010} \\
 H$_2$CO & HITRAN'08\tablenotemark{a} \\
 H$_2$O & \citet{Partridge:1997}; \citet{Gamache:1998}; HITRAN'08\tablenotemark{a} isotopes  \\
 H$_2$S & \citet{Kissel:2002};\citet{Wattson:1992} - private communication; HITRAN'08\tablenotemark{a} isotopes  \\
 HCl & HITRAN'08\tablenotemark{a,d}; \citet{Toth:1970}\\
 HCN & \citet{Harris:2006}; \citet{Harris:2008}; GEISA\tablenotemark{f} isotopes\\
 HF & HITRAN'08\tablenotemark{a,d}; \citet{Mehrotra:1981} \\
 LiCl & \citet{Weck:2004}\tablenotemark{b} \\
 MgH & \citet{Weck:2003b}\tablenotemark{b} \\
 N$_2$ &  HITRAN'08\tablenotemark{a} \\
 NH$_3$ & \citet{Yurchenko:2011}; \citet{Nemtchinov:2004} \\
 NO & HITEMP'10 \tablenotemark{e} \\
 NO$_2$ &  HITRAN'08\tablenotemark{a} \\
 O$_2$ &  HITRAN'08\tablenotemark{a} with 2009 update \\
 O$_3$ & HITRAN'08 \tablenotemark{a,d} \\
 OCS & HITRAN'08\tablenotemark{a} with 2009 update   \\
 OH & \citet{Kurucz:2011}\tablenotemark{g};HITRAN'08\tablenotemark{a} \\
 PH$_3$ &  \citet{Nikitin:2009}; GEISA\tablenotemark{f}; HITRAN'08\tablenotemark{a} \\
 SH &  RLS code \citep{Zare:1973}; \citet{Berdyugina:2002} \\
 SiH & \citet{Kurucz:2011}\tablenotemark{g} \\
 SiO & \citet{Langhoff:1993}; \citet{Kurucz:2011}\tablenotemark{g} \\
 SO$_2$ &  HITRAN'08\tablenotemark{a} with 2009 update \\
 TiO & \citet{Schwenke:1998};\citet{Allard:2000}\\
 VO &  \citet{Alvarez:1998}
\enddata
\tablenotetext{(a)} {\citet{Rothman:2009};\it http://www.cfa.harvard.edu/hitran/updates.html}
\tablenotetext{(b)} {\it http://www.physast.uga.edu/ugamop/}
\tablenotetext{(c)} {\it http://icb.u-bourgogne.fr/OMR/SMA/SHTDS/STDS.html}
\tablenotetext{(d)} {Updates are now available in the HITRAN database. Weaker lines have been added, which would not significantly affect our results.}
\tablenotetext{(e)} {\citet{Rothman:2010};\it http://www.cfa.harvard.edu/hitran/HITEMP.html}
\tablenotetext{(f)} {\it http://ether.ipsl.jussieu.fr/etherTypo/?id=950}
\tablenotetext{(g)} {\it http://kurucz.harvard.edu/molecules.html}
\end{deluxetable}

\begin{deluxetable}{ccccccccc}
\tablecolumns{9}
\tabletypesize{\footnotesize}
\tablewidth{0pt} 
\tablecaption{\label{tab:teff}Effective temperatures.}

\tablehead{\colhead{T$_{\rm surf}$} & 
\multicolumn{4}{c}{BULK SILICATE EARTH}     & 
\multicolumn{4}{c}{CONTINENTAL CRUST} \\
 \colhead{(K)} & 
 \multicolumn{2}{c}{10~m~s$^{-2}$} & 
 \multicolumn{2}{c}{30~m~s$^{-2}$} & 
 \multicolumn{2}{c}{10~m~s$^{-2}$}  & 
 \multicolumn{2}{c}{30~m~s$^{-2}$}\\ 
\colhead{} & 
\colhead{10 bar} & 
\colhead{100 bar} & 
\colhead{10 bar} & 
\colhead{100 bar} & 
\colhead{10 bar} & 
\colhead{100 bar} & 
\colhead{10 bar} & 
\colhead{100 bar}}
\startdata
 1000. & 278. & $-$\tablenotemark{a} & 304.  & $-$\tablenotemark{a} & 355.  & 298. & 389. & 317. \\
 1200. & 323. & 265. & 356.  & 266. & 409.  & 315. & 451. & 340.  \\
 1400. & 373. & 266. & 410.  & 268. & 469.  & 343. & 519. & 375.  \\
 1600. & 413. & 267. & 487.  & 278. & 531.  & 380. & 589. & 419.  \\
 1800. & 519. & 287. & 604.  & 327. & 596.  & 422. & 661. & 467.  \\
 2000. & 647. & 353. & 723.  & 424. & 661.  & 467. & 734. & 515.  \\
 2200. & 766. & 448. & 856.  & 525. & 727.  & 513. & 809. & 569.  \\
\enddata
\tablecomments{The header indicates the composition (Bulk Silicate Earth vs. Continental Crust), and the values for $g$ (in ms$^{-2}$) and $p_{\rm surf}$ (in bar) for each model. The surface temperature is listed in the first column. The other columns list the effective temperature, in K, for the corresponding model. A graphical representation of this table is shown in Figure~\ref{teff}.}
\tablenotetext{(a)} {Not available, since the BSE models with $p_{\rm surf}=100$~bar never reach 1000~K surface temperature. }
\end{deluxetable}

\clearpage

\begin{deluxetable}{cccccccccccccccccc}
\tabletypesize{\scriptsize}
\rotate
\tablecolumns{18}
\tablewidth{0pc}
\tablecaption{\label{magtable}Model Photometry}
\tablehead{Composition & Pressure & Temperature & Gravity & Y & J &	 H &	 K &	 L' &	 M' &	 IRAC1 &	 IRAC2 &	 IRAC3 &	 IRAC4 &	 W1 & W2 & W3& W4 \\
& (bar)& (K) & (m s$^{-2}$)&&&&&&&&&&&&&&}
\startdata
CC & 100 & 1200 & 10 & 33.59 & 32.19 & 32.97 & 31.16	& 28.21 &	22.23 & 30.44 & 22.88 & 24.42 & 22.51 & 31.07 & 22.85 & 20.39 & 18.15	
\enddata
  \label{indexofref}
\end{deluxetable}

\clearpage

\clearpage

\clearpage

\clearpage

\clearpage

\clearpage

\clearpage

\clearpage

\clearpage

\clearpage

\clearpage

\clearpage

\clearpage

\clearpage

\clearpage

\clearpage

\end{document}